\documentclass[superscriptaddress,showpacs,amsmath,amssymb,twocolumn]{revtex4}
\usepackage{amsmath, amsthm, amssymb, bm, braket}
\usepackage[dvips]{graphicx}

\begin{document}

\title{ 
Role of diproton correlation in two-proton emission decay \\
of the $^6$Be nucleus %nuclei beyond the proton drip-line 
}

\author{Tomohiro Oishi}
\author{Kouichi Hagino}
\affiliation{Department of Physics, Tohoku University, Sendai, 980-8578, Japan}
\affiliation{Research Center for Electron Photon Science, Tohoku University, 
1-2-1 Mikamine, Sendai 982-0826, Japan}
%E-mail:oishi@nucl.phys.tohoku.ac.jp
%E-mail:hagino@nucl.phys.tohoku.ac.jp

\author{Hiroyuki Sagawa}
\affiliation{ Center for Mathematics and Physics, University of Aizu, 
Aizu-Wakamatsu, Fukushima 965-8560, Japan}
\affiliation{ RIKEN Nishina Center, Wako 351-0198, Japan}
%E-mail:sagawa@u-aizu.ac.jp

%%%%%%%%%%%%%%%%%%%%%%%%%%%%%%%%%%%%%%%%%%%%%%%%%%%%%%%%%%%%%%%
%%%  You may repeat \author \address as often as necessary  %%%
%%%%%%%%%%%%%%%%%%%%%%%%%%%%%%%%%%%%%%%%%%%%%%%%%%%%%%%%%%%%%%%

\renewcommand{\figurename}{Figure}
\renewcommand{\tablename}{Table}
\renewcommand{\thetable}{\arabic{table}}

\newcommand{\bi}[1]{\ensuremath{\boldsymbol{#1}}}
\newcommand{\unit}[1]{\ensuremath{\mathrm{#1}}}
\newcommand{\oprt}[1]{\ensuremath{\hat{\mathcal{#1}}}}
\newcommand{\abs}[1]{\ensuremath{\left| #1 \right|}}

\def \beq{\begin{equation}}
\def \eeq{\end{equation}}
\def \beqa{\begin{eqnarray}}
\def \eeqa{\end{eqnarray}}
\def \Schr{Schr\"odinger }
\def \twop{2{\it p}}
\def \etal{{\it et al.}}

\def \bir{\bi{r}}
\def \ubir{\bar{\bi{r}}}
\def \bip{\bi{p}}
\def \ubip{\bar{\bi{r}}}

\renewcommand{\include}[1]{} %annul ``include''.
\renewcommand{\documentclass}[2][]{} %annul ``\documentclass''.

\include{begin}
\include{sets}

\begin{abstract}
We discuss a role of diproton correlation 
in two-proton emission from the ground state of a proton-rich nucleus, $^6$Be. 
Assuming the three-body structure of $\alpha + p + p$ configuration, 
we develop a time-dependent approach, in which 
the two-proton emission is described as a time-evolution 
of a three-body metastable state. 
With this method, 
the dynamics of the two-proton emission can be intuitively discussed 
by monitoring the time-dependence of the two-particle density distribution. 
With a model Hamiltonian which well reproduces 
the experimental two-proton decay width, 
we show that 
a strongly correlated diproton emission 
is a dominant process in the early stage of the two-proton emission. 
When the diproton 
correlation is absent, 
the sequential two-proton emission 
competes with the diproton emission, and the decay width is underestimated. 
These results suggest that the two-proton emission decays provide 
a good opportunity to probe the diproton correlation in 
proton-rich nuclei beyond the proton drip-line. 
\end{abstract}

\maketitle

%%%%%%%%%%%%%%%%%%%%%%%%%%%%%%%%%%%%%%%%%%%%%%%%%%%%%%%%%%%%%%%%%%%%%%%%%%%
\section{Introduction} \label{Sec:intro}

The pairing correlation 
plays an essential role in many phenomena 
of atomic nuclei \cite{03Brink,RS80,BCS50,96Doba,03Dean}. 
In recent years, 
the dineutron and diproton correlations 
have particularly attracted a lot of interests 
in connection to the physics of unstable nuclei 
\cite{73Mig,84Catara,91Bert,Zhukov93,OZV99,05Mats,06Mats,
05Hagi,10Ois,10Pillet}. 
These are correlations 
induced by the pairing interaction, with which 
two nucleons are spatially localized. 
Since the pairing gap in infinite nuclear matter 
takes a maximum at the density lower than the normal density 
\cite{06Mats,03Dean,06Cao,07Marg}, 
the dinucleon correlation is enhanced on the surface of nuclei. 
This property may also be related to the BCS-BEC crossover 
\cite{06Mats,07Marg,HSCS07}. 

Although the dinucleon correlation has been theoretically 
predicted for some time, 
it is still an open issue to probe it experimentally. 
For this purpose, 
a pair-transfer reaction 
\cite{91Iga,01Oert,11Shim} and 
the electro-magnetic excitations 
\cite{04Fuku,06Naka,01Myo,07Hagi,07Bertulani_76,11Ois,10Kiku} 
may be considered. 
However, 
even though there have been a few experimental indications \cite{06Naka}, 
so far no direct experimental evidence 
for the dinucleon 
correlation has been found, 
mainly due to a difficulty to access 
the intrinsic structures in bound nuclei 
without disturbing with an external field. 

This difficulty may be overcome by using 
two-proton (\twop-) emission decays (these are referred to as two-proton 
radio-activities when the decay width is sufficiently small) of 
nuclei outside the proton drip-line \cite{08Bla,12Pfu,09Gri_40}. 
An attractive feature of the \twop-emission is that 
two protons are emitted spontaneously from the ground state of unbound 
nuclei, and thus they are expected 
to carry information on the 
pairing correlations inside nuclei, including the 
diproton correlation \cite{05Flam,07Bertulani_34,08Bertulani,12Maru}. 

The \twop-radioactivity was predicted for the first time 
by Goldansky \cite{60Gold,61Gold}. 
He introduced the concept of 
the ``true \twop-decay'', which 
takes place in the situation where 
the emission of single proton is energetically forbidden. 
The pairing interaction plays an important role to generate such 
a situation, lowering the energy of even-Z nuclei. 
In the true \twop-decay process, the two protons may be emitted 
simultaneously as a diproton, that is, the diproton 
decay \cite{60Gold,61Gold,13Deli}. 
This process should thus intimately be related to the diproton correlation. 

Since the time of Goldansky, 
there has been an enormous progress in the problem of 
\twop-decays, both experimentally 
and theoretically, and 
our understanding of the \twop-decays has been 
considerably improved\cite{08Bla,12Pfu,09Gri_40}. 
It has been considered now that 
the actual \twop-decays are often much more complicated 
than the simple diproton decays which Goldansky originally proposed 
\cite{89Boch,05Rotu,07Mie,09Gri_677,09Gri_80,08Muk,10Muk,12Ego,12Gri}. 
Moreover, it has not been completely clarified 
whether the diproton correlation can be actually 
probed by observing \twop-decays. 

The aim of this paper is to investigate the role of the 
diproton correlation in \twop-emissions, 
and discuss a possibility of probing the diproton correlation 
through the \twop-decays. 
For this purpose, one needs to handle 
a many-body meta-stable state, 
for which the theoretical frameworks can be categorized into two approaches: 
the time-independent framework \cite{89Bohm,28Gamov,29Gurney} and 
the time-dependent framework \cite{89Bohm,47Kry,89Kuku}. 
In the time-independent approach, the decay state is 
regarded as a pure outgoing state with a complex energy, that is, 
the Gamow state. 
The real and the imaginary parts of the complex energy 
are related to the decay energy and width, respectively. 
An advantage of this method is that the decay width 
can be accurately calculated even when the width is extremely 
small \cite{09Gri_40,01Gri_I,97Aberg,00Davis}. 
In the time-dependent framework, on the other hand, 
the quantum decay of a metastable state 
is treated as a time-evolution of a wave packet 
\cite{94Serot,98Talou,99Talou_60,00Talou,11Garc,11Campo,12Pons}. 
An advantage of this method is that the decay dynamics can be 
intuitively understood by monitoring the time-evolution of the 
wave packet. 
These two approaches are thus complementary to each other. 

In this paper, we employ the time-dependent approach. 
This approach has been used in Refs. 
\cite{94Serot,98Talou,99Talou_60,00Talou} to study one-proton 
emission decays of proton-rich nuclei. 
In our previous work \cite{12Maru}, 
we extended this approach to \twop-emission in one-dimension. 
We here apply this method to a realistic system, that is, 
the ground state of the $^6$Be nucleus, 
by assuming the three-body structure of $\alpha + p + p$. 
The $^6$Be nucleus is the lightest \twop-emitter, where the \twop-emission 
decay from its ground state has been experimentally studied in Refs. 
\cite{89Boch,09Gri_80,09Gri_677,12Ego}. 
The experimental Q-value of the \twop-emission is 
$1.37$ MeV \cite{88Ajz,02Till}, 
while the $^5$Li nucleus is unbound by 1.96(5) MeV from the threshold 
of $\alpha+p$ \cite{88Ajz}, as shown in Fig. \ref{fig:001}. 
Although the $^5$Li nucleus has a large resonance width of 
about 1.5 MeV \cite{88Ajz,02Till}, 
the $^6$Be nucleus is considered to be a true \twop-emitter. 
Therefore, the sequential decay via the $\alpha + p$ subsystem 
plays a minor role, and 
the effect of the diproton correlation, due to 
the pairing correlation, may significantly be revealed. 
\begin{figure}[tb] \begin{center}
  \fbox {\includegraphics[width=0.9\hsize, clip, trim = 0 0 0 0]
     {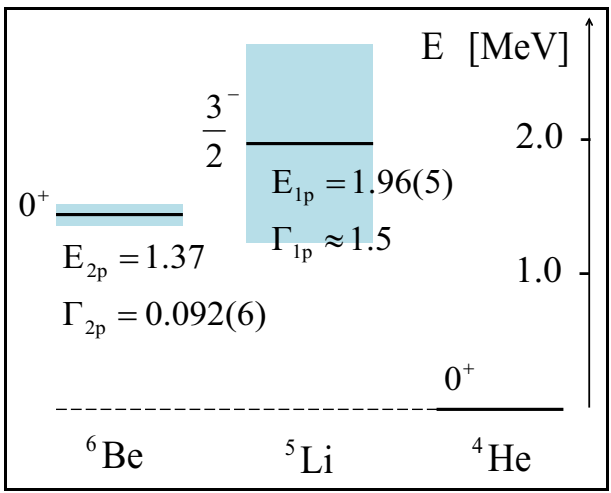}}
  \caption{(Color online) The experimental energy scheme of $^6$Be 
and $^5$Li nuclei with respect to the ground state of $^4$He \cite{88Ajz}. 
The energies and widths are shown in units of MeV. }
\label{fig:001} \vspace{-3truemm}
\end{center} \end{figure}

The paper is organized as follows. 
In Sec. \ref{Sec:form}, we present the theoretical model and 
the time-dependent approach within a quantum three-body model. 
The calculated results for $^6$Be are shown in Sec. \ref{Sec:results}. 
We also discuss the role of pairing 
correlation in the \twop-emission. 
We then summarize the paper in Sec. \ref{Sec:sum}. 

%%%%%%%%%%%%%%%%%%%%%%%%%%%%%%%%%%%%%%%%%%%%%%%%%%%%%%%%%%%%%%%%%%%%%%%%%%%
\section{Formalism} \label{Sec:form}

\subsection{Three-body model Hamiltonian}
In order to describe the \twop-emission from the ground state of 
$^6$Be, we consider a three-body model which 
consists of an $\alpha$-particle 
as the spherical core nucleus and two valence protons. 
As in Refs. \cite{91Bert,05Hagi,10Ois,11Ois}, 
we employ the so called V-coordinate indicated in Fig. \ref{fig:002}. 
Subtracting the center of mass motion of the whole nucleus, 
the total Hamiltonian reads 
\beqa
 H_{\rm 3b} 
 &=& h_1 + h_2 + \frac{\bip_1 \cdot \bip_2}{A_{\rm c} m} + 
   v_{\rm pp}(\bir_1, \bir_2), \\
 h_i 
 &=& \frac{\bip_i^2}{2\mu} + V_{\rm cp}(r_i) \; \; \; \; (i=1,2), 
\eeqa
where $h_i$ is the single particle (s.p.) Hamiltonian between 
the core and the $i$-th proton. 
$\mu \equiv m A_{\rm c} / (A_{\rm c}+1)$ is the reduced mass 
where $m$ and $A_{\rm c}$ are the nucleon mass and the mass 
number of the core nucleus, respectively. 
The interaction between $\alpha$ and a valence proton, $V_{\rm cp}$, 
consists of 
the nuclear potential $V_{\rm WS}$ and 
the Coulomb potential $V_{\rm Coul}$, 
\beq
 V_{\rm cp}(r_i) = V_{\rm WS}(r_i) + V_{\rm Coul}(r_i). 
\eeq
For the Coulomb part of the potential, $V_{\rm Coul}$, 
we use the one appropriate to a uniformly charged
spherical alpha particle of radius $r_c=1.68$ fm. 
For the nuclear part, $V_{\rm WS}$, 
on the other hand, we use the Woods-Saxon parametrization given by 
\beq
 V_{\rm WS}(r) = 
 V_0 f(r) + U_{ls} (\bi{l} \cdot \bi{s}) \frac{1}{r} \frac{df(r)}{dr}, 
\label{eq:cp-WS}
\eeq
where 
\beq
 f(r) = \frac{1}{1 + e^{(r-R_0)/a_0} } 
\eeq
with $R_0 = r_c$ and $a_0$ = 0.615 fm. 
We use the depth parameters of $V_0 = -58.7$ MeV and 
$U_{ls} = 46.3$ ${\rm MeV\cdot fm^2}$. 
This potential yields 
the resonance energy and the width 
of the $(p_{3/2})$-channel for 
$\alpha-p$ scattering of 
$E_r (p_{3/2}) = 1.96$ MeV and 
$\Gamma_r (p_{3/2}) = 1.56$ MeV, respectively. 
These values are compared with the experimental data, 
$E_r (p_{3/2}) = 1.96(5)$ MeV and 
$\Gamma_r (p_{3/2}) \sim 1.5$ MeV (see Fig.\ref{fig:001}) \cite{88Ajz}. 
We note that this resonance state is quite broad 
and there has been some ambiguity in the observed 
decay width \cite{88Ajz,02Till,00Hoef,09Shir}. 
\begin{figure}[tb] \begin{center}
 \fbox{ \includegraphics[width=0.8\hsize, clip, trim = 10 0 10 10]
    {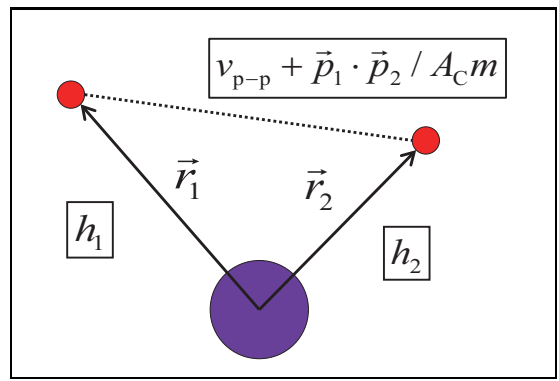}}
  \caption{(Color online) The V-coordinate for three-body system. }
\label{fig:002} \vspace{-3truemm}
\end{center} \end{figure}

For the proton-proton interaction, $v_{\rm pp}$, 
we use the Minnesota potential \cite{77Thom} 
together with the Coulomb term for point charges: 
\beq
 v_{\rm pp}(\bir_1, \bir_2) = 
 v_0 e^{-b_0 r_{12}^2} + v_1 e^{-b_1 r_{12}^2} % v_{\rm pp}^{(N)}(r_{12}) 
 + \frac{e^2}{r_{12}}, %\frac{e^2}{4\pi \epsilon_0} 
\label{eq:Minne} 
\eeq
where $r_{12} = \abs{\bir_1 - \bir_2}$. 
For $b_0, b_1$ and $v_1$, 
we use the same parameters introduced in the original paper \cite{77Thom}, 
as summarized in Table \ref{tb:3}. 
%that is, $b_0$ = 1.48 fm$^{-2}$, $b_1$ =0.639 fm$^{-2}$, and 
%$v_1=-178.0$ MeV, and 
%$b_0$ = 1.48 fm$^{-2}$, $b_1$ =0.465 fm$^{-2}$, and 
%$v_1=-91.85$ MeV for 
%$S=0$ and $S=1$, respectively, for 
%the combined spin of the two protons. 
On the other hand, the strength of the repulsive term, $v_0$, 
is adjusted so as to reproduce the empirical Q-value 
for the two-proton emission, as we will discuss in Sec. \ref{Sec:results}. 
\begin{table}[tb] \begin{center}
  \catcode`? = \active \def?{\phantom{0}} %define `?' as ' '(one-blank).
  \begingroup \renewcommand{\arraystretch}{1.2}
  \begin{tabular*}{\hsize} { @{\extracolsep{\fill}} cccc} \hline \hline
          & $b_0$ (fm$^{-2}$) & $b_1$ (fm$^{-2}$) & $v_1$ (MeV) \\ \hline
    $S=0$ & 1.48             & 0.639           & $-178.0$    \\
    $S=1$ & 1.48             & 0.465           & $-91.85$    \\ \hline \hline
  \end{tabular*}
  \endgroup
  \catcode`? = 12 %initialize `?'.
  \caption{The values of $b_0,b_1$ and $v_1$ in the Minnesota 
potential given by Eq.(\ref{eq:Minne}). 
$S$ indicates the combined spin of the two protons. } 
\label{tb:3} \vspace{-3truemm}
\end{center} \end{table}

\subsection{Uncorrelated Two-proton Basis}
Each s.p. state satisfying 
$h_i \phi_a (\bir_i) = \epsilon_a \phi_a (\bir_i)$ is 
labeled by $a=\left\{ n_a, l_a, j_a, m_a \right\}$, that is, 
a combination of the radial quantum 
number $n$, the orbital angular momentum $l$, the spin-coupled angular 
momentum $j$ and its z-component $m$. 
%We assume that the $\alpha$-p potential is spherical, and independent of 
%spin variables. 
%Thus we only solved the radial part of s.p. wave functions numerically with 
%Numerov's method \cite{93Hairer}. 
Using these s.p. wave functions, 
one can construct the uncorrelated basis for the two protons 
coupled to an arbitrary spin-parity, $J^{\pi}$, where 
the coupled angular momentum $J$ is given by $J = j_a \oplus j_b$ 
and the total parity $\pi$ is given by $\pi = (-)^{l_a+l_b}$. 
That is, 
\beq
 \Phi_{ab}^{(J^{\pi})} (\bir_1, \bir_2) 
 = \oprt{A} \left[ \phi_a (\bir_1) \otimes 
                   \phi_b (\bir_2) \right]^{(J^\pi)}, 
 \label{eq:basis}
\eeq
where $\oprt{A}$ is the anti-symmetrization operator. 
In this work, we assume that the core nucleus always stays in the 
ground state with the spin-parity of $0^+$. 
Thus the uncorrelated basis given by Eq.(\ref{eq:basis}) are reduced only 
to the $J^\pi=0^+$ subspace, 
since the ground state of $^6$Be also has the spin-parity of $0^+$. 
That is, 
\beqa
 & & \Phi_{ab}^{(0^+)} (\bir_1, \bir_2) 
     = \Phi_{n_a n_b lj} (\bir_1, \bir_2) \\
 & & \phantom{00} = \frac{1}{\sqrt{2(1+\delta_{n_a,n_b})} } \sum_m C(j,m;j,-m \mid 0,0) \nonumber \\
 & & \phantom{0000} \left[   \phi_{n_a ljm} (\bir_1) \phi_{n_b lj-m} (\bir_2) \right. \nonumber \\
 & & \phantom{0000} \left. + \phi_{n_a ljm} (\bir_2) \phi_{n_b lj-m} (\bir_1) \right] . 
 \label{eq:basis_0p} 
\eeqa
Notice $l_a = l_b$, $j_a =j_b$ for the $0^+$ state. 
In the following, for simplicity, 
we omit the superscript $(0^+)$ and 
use a simplified notation, $\ket{\Phi_M}$, 
for the uncorrelated basis given by Eq. (\ref{eq:basis_0p}), 
where $M=(n_a, n_b, l,j)$. 

The eigen-states of the three-body Hamiltonian, $H_{\rm 3b}$, can be 
obtained by expanding the wave function on the uncorrelated basis, 
\beq
 \ket{E_N} = \sum_M U_{NM} \ket{\Phi_M}, 
\label{eq:expansion} 
\eeq
where the expansion coefficients, $U_{NM}$, are determined by 
diagonalizing the Hamiltonian matrix for $H_{\rm 3b}$. 
The state $\ket{E_N}$ then satisfies 
$H_{\rm 3b} \ket{E_N} = E_N \ket{E_N}$ in a truncated space. 

All our calculations are performed in the truncated space 
defined by the energy-cutoff, 
$\epsilon_a + \epsilon_b \leq E_{\rm cut} =40$ MeV. 
The continuum s.p. states are discretized within the radial box of 
$R_{\rm box}=80$ fm (notice that the states $\ket{E_N}$ are also 
discretized). 
%We choose relatively a compact box with lower costs of calculations, 
%for the first testing purpose of our method. 
For the angular momentum channels, we include from $(s_{1/2})^2$ to 
$(h_{11/2})^2$ configurations. 
In order to take into account the effect of the Pauli principle, 
we exclude the bound 1$s_{1/2}$ state from Eq.(\ref{eq:expansion}), 
that is given by the protons in the core nucleus. 
We have confirmed that our conclusions do not change 
even if we employ a larger value of $E_{\rm cut}$ and/or 
include higher partial waves. 

\subsection{Time-Dependent Method for Two-Proton Decay}
Assuming the \twop-emission as a time-dependent process, 
we carry out time-dependent calculations for 
the three-body system, $^6$Be. 
For this purpose, we first need to determine 
the initial state, $\ket{\Psi (t=0)}$, 
for which the 
two valence protons are confined inside the 
potential barrier generated by the core nucleus. 
That is, 
the \twop-density distribution at $t=0$ has almost no amplitude 
outside the potential barrier. 
In order to construct such initial state, 
we employ the confining potential method, which will be detailed 
in the next section. 

The initial state so obtained can be expanded with the eigen-states 
of $H_{\rm 3b}$, that is, $\ket{E_N}$ given by Eq. (\ref{eq:expansion}) as
\beq
 \ket{\Psi (0)} = \sum_N F_N (0) \ket{E_N}. \label{eq:expand_E} 
\eeq
After the time-evolution with the three-body Hamiltonian $H_{\rm 3b}$, 
this state is evolved to 
\beq
 \ket{\Psi (t)} 
 =  \exp \left[ -it \frac{H_{\rm 3b}}{\hbar} \right] 
           \ket{\Psi (0)} = \sum_{N} F_N (t) \ket{E_N}, \label{eq:ex_E} 
\eeq
where 
\beq
 F_N (t) = e^{-itE_N/\hbar} F_N(0). \label{eq:excf_E}
\eeq
Notice that the state $\ket{\Psi (t)}$ 
can also be expanded on 
the uncorrelated basis as, 
\beq
 \ket{\Psi (t)} 
 = \sum_{M} C_M (t) \ket{\Phi_M}, 
 \label{eq:tdse}
\eeq
with 
\beq
 C_M (t) = \sum_N F_N (t) U_{NM}. \label{eq:excf_UNCB} 
\eeq

We define 
the Q-value of the \twop-emission as the expectation 
value of the total Hamiltonian, $H_{\rm 3b}$, with 
respect to the initial state, $\ket{\Psi(0)}$. 
Since the time-evolution operator, 
$\exp \left[ -it H_{\rm 3b}/\hbar \right]$, in Eq. (\ref{eq:ex_E}) 
commutes with $H_{\rm 3b}$, 
the Q-value is conserved during the time-evolution. 
That is, 
\beq
 Q  = \Braket{\Psi (0) | H_{\rm 3b} | \Psi (0)} 
    = \Braket{\Psi (t) | H_{\rm 3b} | \Psi (t)}. \label{eq:Q_value}
\eeq
We also note that the wave function is normalized 
at any time: $\Braket{\Psi(t) | \Psi(t)} = 1$. 

In order to extract the information on the dynamics 
of two-proton emission, 
it is useful to introduce the decay state, 
$\ket{\Psi_d (t)}$, which is defined as the orthogonal 
component of 
$\ket{\Psi(t)}$ 
to 
the initial state \cite{08Bertulani}. 
That is, 
\beq
 \ket{\Psi_d (t)} \equiv 
 \ket{\Psi (t)} - \beta(t) \ket{\Psi (0)}, \label{eq:decaystate}
\eeq
where $\beta (t) = \Braket{\Psi (0) | \Psi (t)}$. 
While the initial state 
is almost confined inside the potential barrier, 
the main part of the decay state is located outside the barrier. 
We define 
the decay probability as the norm of the decay state, 
\beq
 N_d (t) 
 \equiv \Braket{\Psi_d (t) | \Psi_d (t)} 
 = 1 - \abs{\beta (t)}^2 \label{eq:603DComp}. 
\eeq
Notice that $N_d(0) = 0$ since $\beta (0) = 1$. 
Because $\abs{\beta (t)}^2$ is identical to 
the survival probability for the decaying process, 
the decay width can be defined with $N_d (t)$ 
as \cite{94Serot, 98Talou, 99Talou_60, 00Talou}, 
\beq
 \Gamma (t) 
 \equiv  -\hbar \frac{d}{dt} \ln \left[ 1-N_d(t) \right] 
 = \frac{\hbar }{1-N_d(t)} \frac{d}{dt} N_d(t). \label{eq:width} 
\eeq
It is worthwhile to mention that 
if the time-evolution 
follows the exponential decay law, 
such that 
\beq
 \left[ 1-N_d(t) \right] = e^{-t/\tau}, 
\eeq
then $\Gamma(t)$ is related to the lifetime of the 
meta-stable state: $\Gamma = \hbar / \tau$. 
This situation is 
realized when 
the energy spectrum, 
defined by $\{ \abs{F_N(t)}^2 \}$, 
is well approximated as a Breit-Wigner 
distribution \cite{47Kry, 89Kuku}. 

It is useful to define also the partial decay width $\Gamma_s(t)$ 
in order to understand the decay dynamics. 
This is defined as the width for the decay to a channel $s$, where the total 
decay width is given by 
\beq
 \Gamma (t) = \sum_s \Gamma_s (t). 
\eeq
The partial decay width can be calculated with the expansion 
coefficient $a_s(t)$ 
of the decay state with the channel wave function, 
\beq
 \ket{\Psi_d (t)} = \sum_s a_s(t) \ket{s}, 
 \label{eq:part_expand} 
\eeq
where $\Braket{s' | s} = \delta_{s's}$. 
Since $N_d(t)$ in Eq. (\ref{eq:603DComp}) is given as 
$N_d(t)=\sum_s|a_s(t)|^2$, 
the partial decay width reads 
\beq
 \Gamma_s (t) = \frac{\hbar }{1-N_d(t)} \frac{d}{dt} N_{d,s}(t). 
 \label{eq:pwidth} 
\eeq
where $N_{d,s} = \abs{a_s(t)}^2$. 
In the next section, we will apply 
Eq. (\ref{eq:pwidth}) in order to 
calculate the spin-singlet and spin-triplet widths 
for the \twop-emission of $^6$Be. 

%We note that Eqs.(\ref{eq:part_expand}) and (\ref{eq:pwidth}) 
%can be formulated generally for any choice of partial components 
%as long as those are orthogonal. 
%For example, one can employ the components which have different 
%energies, angular momenta or spin-parities. 
%We also emphasize that our formulas themselves in this subsection 
%are not limited in the three-body framework, and can be expanded to 
%further complex models. 

%%%%%%%%%%%%%%%%%%%%%%%%%%%%%%%%%%%%%%%%%%%%%%%%%%%%%%%%%%%%%%%%%%%%%%%%%%%
\section{Results} \label{Sec:results}

\subsection{Initial State}
Let us now numerically solve the three-body model and discuss the 
\twop-decay of $^6$Be. 
As we mentioned in the previous section, 
we construct the initial state for the two protons 
such that the \twop-density distribution is localized around 
the core nucleus and thus 
has almost no amplitude outside the core-proton potential barrier. 
To this end, we employ 
the confining potential method \cite{87Gurv, 88Gurv, 04Gurv}. 
Within this method, 
we modify the core-proton potential, $V_{\rm cp}$, 
so as to make a meta-stable two-proton state be bound. 
\begin{figure}[t] \begin{center}
  \fbox{ \includegraphics[width=0.8\hsize, clip, trim = 10 0 5 5]{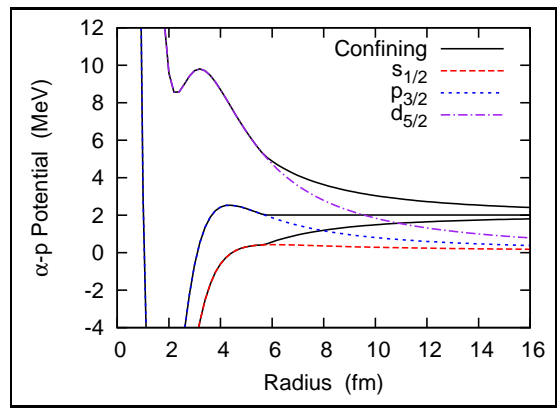}}
  \caption{(Color online) The original and confining potentials for 
the $s_{1/2}$, $p_{3/2}$ and $d_{5/2}$ channels 
in the $\alpha-p$ subsystem. 
$R_b$ in Eqs. (\ref{eq:Vconf}) and (\ref{eq:Vconf2}) is taken to be 
5.7 fm for all the channels. } \label{fig:1} \vspace{-3truemm}
\end{center} \end{figure}

\begin{figure*}[htb] \begin{center}
  \begin{tabular}{c} %switch-off the auto-turning
     $^6$Be ($ct=0$) \\
     \begin{minipage}{0.5\hsize} \begin{center}
     \fbox{ \includegraphics[height=45truemm, clip, trim = 20 0 10 25]
            {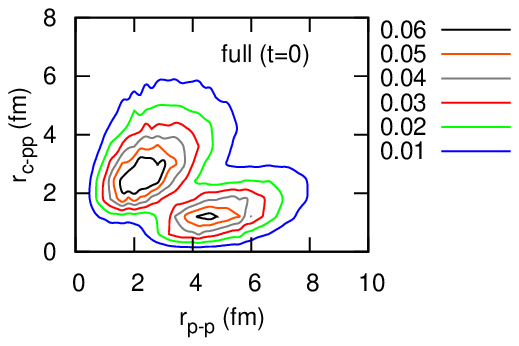}}
     \end{center} \end{minipage}
     \begin{minipage}{0.5\hsize} \begin{center}
     \fbox{ \includegraphics[height=45truemm, clip, trim = 10 5 5 5]
            {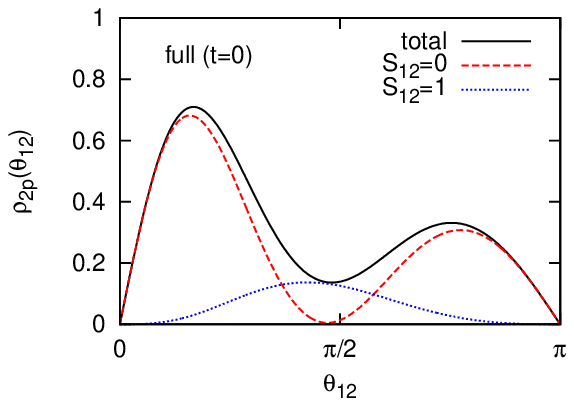}}
     \end{center} \end{minipage}
  \end{tabular}
  \caption{(Color online) The left panel: 
The \twop-density at $t=0$ 
obtained by including all the configurations up to $(h_{11/2})^2$. 
It is plotted as a function of 
$r_{\rm p-p} = (r_1^2+r_2^2-2r_1r_2\cos \theta_{12})^{1/2}$ and 
$r_{\rm c-pp} = (r_1^2 + r_2^2 + 2r_1r_2\cos \theta_{12})^{1/2}/2$. 
The right panel: 
The angular distributions at $t=0$ obtained by integrating 
$\bar{\rho}_{2p}$ with respect to 
$r_1$ and $r_2$. } \label{fig:2} \vspace{-3truemm}
\end{center} \end{figure*}

\begin{figure*}[htb] \begin{center}
  \begin{tabular}{c} %switch-off the auto-turning
     $^6$Be ($ct=0$), $(l=odd)^2$ only \\
     \begin{minipage}{0.5\hsize} \begin{center}
       \fbox{ \includegraphics[height=45truemm, clip, trim = 20 0 10 25]
              {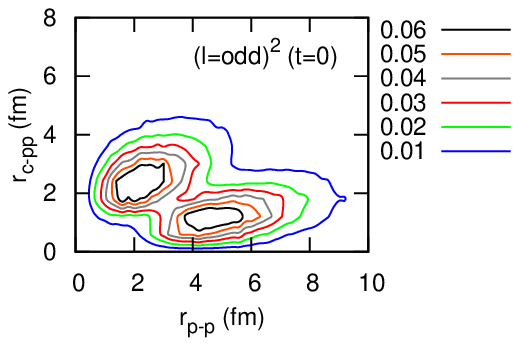}}
     \end{center} \end{minipage}
     \begin{minipage}{0.5\hsize} \begin{center}
     \fbox{ \includegraphics[height=45truemm, clip, trim = 10 5 5 5]
              {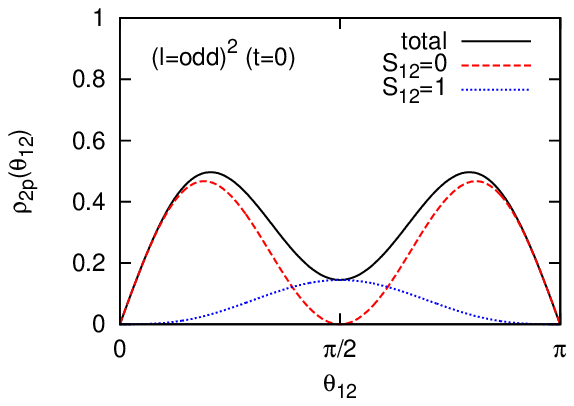}}
     \end{center} \end{minipage}
  \end{tabular}
  \caption{(Color online) The same as Fig. \ref{fig:2} but in the case 
with only odd-$l$ partial waves. } \label{fig:3} \vspace{-3truemm}
\end{center} \end{figure*}

We generate the confining potential for the 
present problem as follows. 
Because the $\alpha$-p subsystem has a resonance at 
$E_0=1.96$ MeV in the $p_{3/2}$-channel, 
the two protons in $^6$Be are expected to have a large component of 
the $(p_{3/2})^2$ configuration. 
Thus, we first modify the core-proton potential for the 
$p_{3/2}$-channel in order to 
generate a bound state as follows: 
\beqa
 V_{{\rm cp}}^{conf}(r) \nonumber 
 &=& V_{{\rm cp}}(r) \phantom{000} (r \leq R_b), \\
 &=& V_{{\rm cp}}(R_b) \phantom{00} (r > R_b), 
\label{eq:Vconf}
\eeqa
with $R_b=5.7$ fm. 
Here we have followed Ref. \cite{04Gurv} and taken $R_b$ to be 
outside the potential barrier rather than the barrier position. 
For the other s.p. channels, we define the confining potential as 
\beq
 V_{\rm cp}^{conf}(r) %\nonumber 
 =\left\{ \begin{array}{cc} 
      V_{\rm cp}(r) \phantom{0000} & (r \leq R_b), \\
      V_{\rm cp}(r) + \Delta V_{p_{3/2}}(r) & (r > R_b), \end{array} \right.
\label{eq:Vconf2}
\eeq
%where $V_b(r) = V_{{\rm cp},~(p_{3/2})}(R_b) - V_{{\rm cp},~(p_{3/2})}(r)$. 
where $\Delta V_{p_{3/2}}(r) = V_{{\rm cp}}(R_b) - V_{{\rm cp}}(r)$ for the 
$p_{3/2}$-channel. 
The original and confining potentials for the 
$s_{1/2}$, $p_{3/2}$ and $d_{5/2}$ channels are shown in 
Fig. \ref{fig:1}. 
We note that, for this system, the core-proton barrier 
is mainly due to the centrifugal potential 
rather than the Coulomb potential. 
This situation is quite different from heavy \twop-emitters 
with a large proton-number, such as $^{45}$Fe.
 \begin{figure*}[htb] \begin{center}
  $^{6}$He (g.s.) \\
  \begin{tabular}{c} %switch-off the auto-turning
     \begin{minipage}{0.5\hsize} \begin{center}
        \fbox{ \includegraphics[height=45truemm, clip, trim = 20 0 10 20]
               {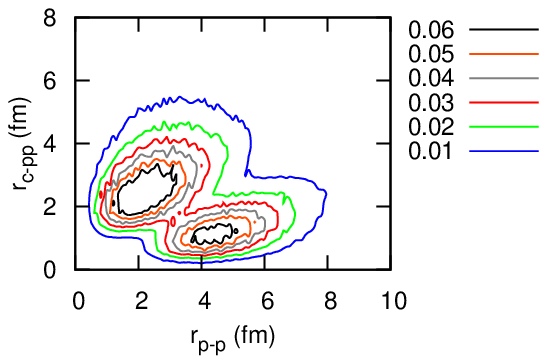}}
     \end{center} \end{minipage}
     \begin{minipage}{0.5\hsize} \begin{center}
        \fbox{ \includegraphics[height=45truemm, clip, trim = 10 5 5 5]
               {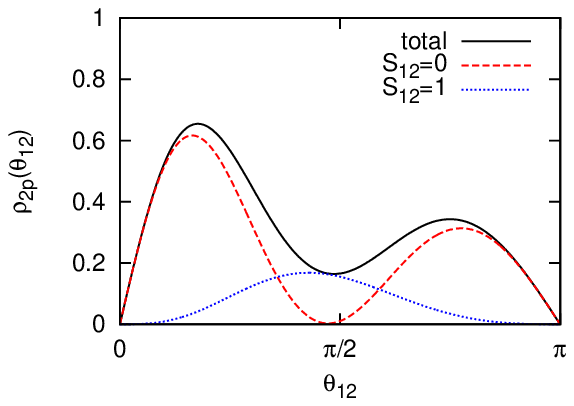}}
     \end{center} \end{minipage}
  \end{tabular}
  \caption{(Color online) The density distribution of 
the valence two neutrons, $\bar{\rho}_{2n}$, in the ground state of $^6$He. 
Those are plotted in the same manner as in the left and right panels 
of Fig. \ref{fig:2}. 
The configurations up to $(h_{11/2})^2$ are included. } \label{fig:22}
\end{center} \end{figure*}

The initial state for the \twop-emission 
is obtained by diagonalizing the modified 
Hamiltonian including $V^{conf}_{\rm cp}(r)$. 
The empirical Q-value for the two-proton emission is 1.37 MeV 
for $^6$Be \cite{88Ajz, 02Till}. 
However, the Minnesota potential with the 
original parameters 
overestimates this value by about 50\%. 
Thus we have modified the parameter $v_0$ in Eq.(\ref{eq:Minne}) from the 
original value, $v_0 = 200.0$ MeV \cite{77Thom}, 
to $v_0 = 156.0$ MeV so as to yield $Q=1.37$ MeV 
when it is calculated by Eq.(\ref{eq:Q_value}). 

In Fig. \ref{fig:2}, we show the density distribution of the 
initial state obtained in this way. 
By integrating the spin coordinates, 
the density distribution becomes a function of the radial 
distances, $r_1$ and $r_2$, as well as the opening angle between 
the two valence protons, $\theta_{12}$. 
That is, 
\beqa
 & & \bar{\rho}_{2p} (t=0; r_1,r_2,\theta_{12}) \nonumber \\
 & & \phantom{000} \equiv 8\pi^2 r_1^2 r_2^2 \sin \theta_{12}
     \cdot \rho_{2p} (t=0; r_1,r_2,\theta_{12}), 
\eeqa
with 
\beq
 \rho_{2p}(t=0; r_1,r_2,\theta_{12}) = \abs{\Psi(t=0; r_1,r_2,\theta_{12})}^2. 
\eeq
Here $\bar{\rho}_{2p}$ is normalized as 
\beq
 \int_0^{R_{\rm box}} dr_1 \int_0^{R_{\rm box}} dr_2 
 \int_{0}^{\pi} d\theta_{12} \,\bar{\rho}_{2p} = 1. 
\eeq
In the left panel of Fig. \ref{fig:2}, $\bar{\rho}_{2p}$ is plotted as a 
function of the distance between the core and the center of 
mass of the two protons: 
$r_{\rm c-pp} = \sqrt{r_1^2 + r_2^2 + 2r_1r_2\cos \theta_{12}}/2$, 
and the relative distance between the two protons: 
$r_{\rm p-p} = \sqrt{r_1^2 + r_2^2 - 2r_1r_2\cos \theta_{12}}$. 
In the right panel of Fig. \ref{fig:2}, we also display 
the angular distributions obtained by integrating $\bar{\rho}_{2p}$ 
for the radial distances. 

It is clearly seen that the initial wave function 
is confined inside the potential barrier at $r \cong 4$ fm 
(see Fig. \ref{fig:1}). 
Furthermore, the \twop-density is concentrated 
near $r_{\rm p-p} = 2$ fm, corresponding to the 
diproton correlation in bound nuclei \cite{10Ois}. 
The corresponding angular distribution becomes asymmetric 
and has the higher peak at the opening angle $\theta_{12} \cong \pi /6$. 
This peak is almost due to the spin-singlet configuration, 
being analogous to the dinucleon correlation. 
This suggests the existence of 
the diproton correlation in the 
meta-stable ground state of $^6$Be due to the pairing correlation. 

As is well known, the mixture of configurations with s.p. states 
with opposite parity 
plays an essential role in generating 
the dinucleon correlation \cite{84Catara}. 
In order to study the effect of the diproton correlation 
in the \twop-emission, 
we have also performed the same calculation but only with 
odd-$l$ partial waves, that is, $p^2, f^2$, and $h^2$. 
In the following, we call this case as the $(l=odd)^2$ case. 
In this case, the pairing correlations are %% HS %%partially
taken into account only among the s.p. states with the same parity, 
%%  HS  %%although
while the mixture of opposite parity configurations 
is entirely ignored. 
In Fig. \ref{fig:3}, we show the initial configuration obtained only 
with the odd-$l$ partial waves. 
We have used $v_0 = 88.98$ MeV in order to reproduce 
the empirical Q-value in this case. 
In the left panel of Fig. \ref{fig:3}, 
there are two comparable peaks at $r_{\rm p-p} = 2$ and $5$ fm 
whereas, in the right panel, 
the corresponding angular distribution has a symmetric form. 
This result is in contrast with that in the case with all the 
configurations from $(s_{1/2})^2$ to $(h_{11/2})^2$, 
shown in Fig. \ref{fig:2}, 
where the pairing correlations are fully taken into account. 
\begin{table}[b] \begin{center}
  \catcode`? = \active \def?{\phantom{0}} %define `?' as ' '(one-blank).
  \begingroup \renewcommand{\arraystretch}{1.2}
  \begin{tabular*} {\hsize} { @{\extracolsep{\fill}} ccccc cc} \hline \hline
  && \multicolumn{2}{c}{$^{6}$Be ($t=0$)} && $^{6}$He (g.s.) & \\ \cline{3-4} \cline{6-6}
  && full & $(l=odd)^2$ && full    & \\ \hline
   $\Braket{H_{\rm 3b}}$ (MeV) && 1.37 & 1.37 && $-0.975$ & \\
  &&&&& & \\
   $(p_{3/2})^2$ (\%)        && 88.9  & 97.1 && 92.7 & \\
   $(p_{1/2})^2$ (\%)        && ?3.1  & ?2.8 && ?1.6 & \\
   $(s_{1/2})^2$ (\%)        && ?2.2  & ?0.0 && ?1.3 & \\
   other $(l=even)^2$ (\%) && ?5.2  & ?0.0 && ?4.2 & \\
   other $(l=odd)^2$ (\%)  && ?0.6  & ?0.1 && ?0.2 & \\
  &&&&& & \\
   $P(S=0)$ (\%) && 82.2 & 80.6 && 78.1 & \\ \hline \hline
  \end{tabular*}
  \endgroup
  \catcode`? = 12 %initialize `?'.
  \caption{Calculated properties for the initial state of $^6$Be 
and the bound ground state of $^6$He. 
The results with all the configurations 
from $(s_{1/2})^2$ to $(h_{11/2})^2$ are labeled by ``full''. 
Those obtained only with the odd-$l$ partial waves for 
$^6$Be are also shown. } \label{tb:1}
\end{center} \end{table}

In Table \ref{tb:1}, properties of the initial state are summarized. 
It is clearly seen that, 
in the case of the full configuration-mixture, 
the main component is $(p_{3/2})^2$, 
reflecting the fact that the $p_{3/2}$ channel has a 
resonance in the $\alpha$-p subsystem. 
The mixture of different partial waves are due to the 
off-diagonal matrix elements of $H_{\rm 3b}$, 
corresponding to the pairing correlations. 
A comparable enhancement of the spin-singlet configuration exists 
also in the case with the $(l=odd)^2$ bases, 
even though there is no localization of the two protons 
as shown in Fig. \ref{fig:3}. 

%\subsubsection{Comparison with $^6$He}
From the point of view of the isobaric symmetry in nuclei, 
it is interesting to compare the initial state of $^6$Be with the 
ground state of its mirror nucleus, $^6$He. 
Assuming the $\alpha$+n+n structure, 
we perform the similar calculation for the ground state of $^6$He. 
For the $\alpha$-n system, there is an observed 
resonance of $p_{3/2}$ at 
$E_r=0.735(20)$ MeV with its width, 
$\Gamma_r=0.600(20)$ MeV \cite{88Ajz,NNDCHP}. 
In order to reproduce this resonance, we exclude the Coulomb term 
from $V_{\rm cp}$ and modify the depth parameter 
to $V_0 = -61.25$ MeV in the Woods-Saxon potential. 
The pairing interaction is adjusted to reproduce 
the empirical two-neutron separation energy, 
$\Braket{H_{\rm 3b}}=-S_{\rm 2n}=-0.975$ MeV \cite{NNDCHP}, 
by using $v_0 = 212.2$ MeV in Eq.(\ref{eq:Minne}). 
Notice that we deal with the bound state of the three-body 
system in this case, and thus the confining potential is 
not necessary. 
%One may concern the difference of $v_0$ between $^6$Be and $^6$He. 
%This might be due to an ambiguity in $V_{\rm cp}$ for $^6$Be 
%originated from a broad resonance in the core-proton subsystem. 
%Improving $V_{\rm cp}$ in $^6$Be can lead to the more consistent 
%set of parameters among $V_{\rm c-N}$ and $v_{\rm N-N}$. 
%We note that this ambiguity does not affect our qualitative discussions. 
In Fig. \ref{fig:22}, the two-neutron density distribution is 
shown in the same manner as in Fig. \ref{fig:2}. 
Its properties are also summarized in the last column of Table \ref{tb:1}. 
Obviously, the two-neutron wave function in $^6$He 
has a similar distribution to 
the \twop-wave function in $^6$Be. 
The dinucleon correlation is present also in $^6$He, 
characterized as the spatial localization with the 
enhanced spin-singlet component \cite{05Hagi}. 
Consequently, the confining potential which we employ 
provides such initial state of $^6$Be 
that can be interpreted as the isobaric analogue state of $^6$He. 
\begin{figure}[t] \begin{center}
 \fbox{ \includegraphics[width=0.8\hsize, clip, trim = 0 0 0 0]
        {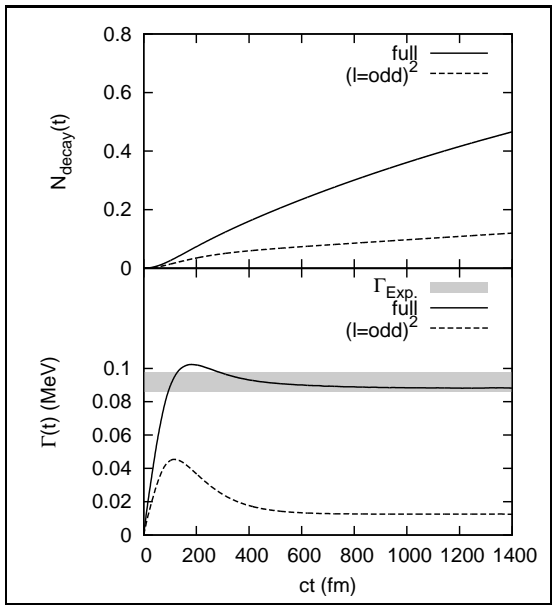}}
 \caption{The decay probability and the decay width 
of the \twop-emission from $^6$Be, obtained with the time-dependent method. 
The result in the case of the full configuration-mixture 
is plotted by the solid line, 
whereas that in the $(l=odd)^2$ case is plotted by the dashed line. 
The experimental value, $\Gamma_{\rm exp} = 92 \pm 6$ keV \cite{88Ajz,02Till}, 
is marked by the shaded area. 
} \label{fig:4} \vspace{-3truemm}
\end{center} \end{figure}

\subsection{Decay Width}
In order to describe the decay process 
of $^{6}$Be, we suddenly change the potential at $t=0$ from the 
confining potential, $V_{\rm cp}^{conf}$, to the original one, 
$V_{\rm cp}$. 
The initial state constructed in the previous subsection then 
evolves in time. 
We first show the results of the decay probability, $N_d(t)$, 
and the decay width, $\Gamma (t)$, defined by Eqs.(\ref{eq:603DComp}) 
and (\ref{eq:width}), respectively. 

In Fig. \ref{fig:4}, the calculation is carried out up to 
$ct = 1400$ fm. 
We have confirmed that the artifact due to the reflection 
at $r=R_{\rm box}$ is negligible in this time-interval. 
One can clearly see that, 
after a sufficient time-evolution, the decay width converges 
to a constant value for all the cases, 
and the exponential decay-rule is realized. 
Furthermore, the result in the case of full configuration-mixture 
yields the saturated value of $\Gamma (t) \cong 88.2$ keV, which 
reproduces the experimental decay width, 
$\Gamma = 92 \pm 6$ keV \cite{88Ajz,02Till}. 

On the other hand, 
the decay width is significantly underestimated 
when the partial waves are limited only to odd-$l$ partial waves 
(Note that we exclude even-$l$ partial waves not only at $t=0$ but 
also for $t>0$ in this case). 
The underestimation of the decay width 
is caused by an increase of the pairing attraction: 
with the odd-$l$ partial waves only, to reproduce the empirical Q-value, 
we needed a stronger pairing attraction. 
The two protons are then strongly bound to each other 
and are difficult to go outside, 
even they have a similar energy release to that 
in the case of full configuration-mixture. 
From this result, we can conclude that the mixing of opposite 
parity configurations is indispensable in order to 
reproduce simultaneously the Q-value and the decay 
width of the \twop-emission, 
supporting the assumption of the diproton correlation. 
\begin{figure}[t] \begin{center}
 \fbox{ \includegraphics[width=0.8\hsize, clip, trim = 10 0 0 5]
        {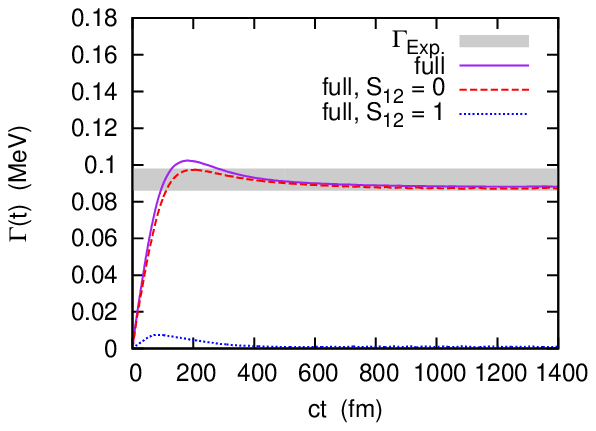}}
 \fbox{ \includegraphics[width=0.8\hsize, clip, trim = 10 0 0 5]
        {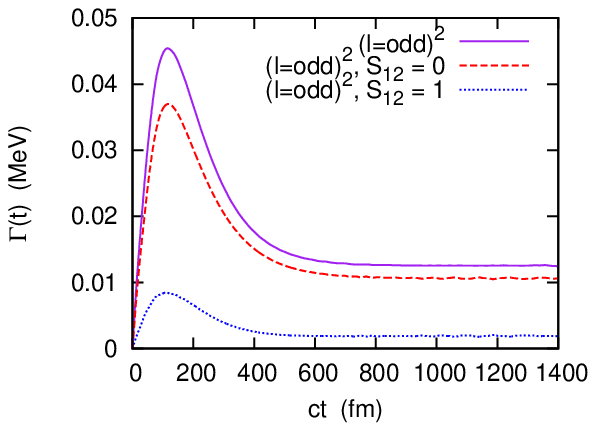}}
 \caption{(Color online) The partial decay widths 
of the spin-singlet and the spin-triplet configurations 
in the \twop-emission of $^6$Be. 
In the upper panel, the result obtained with all the 
configurations from $(s_{1/2})^2$ to $(h_{11/2})^2$ is shown. 
In the lower panel, the same result but in the case 
with only odd-$l$ partial waves is plotted. } 
\label{fig:5-1} \vspace{-3truemm}
\end{center} \end{figure}
\begin{table}[tb] \begin{center}
  \catcode`? = \active \def?{\phantom{0}} %define `?' as ' '(one-blank).
  \begingroup \renewcommand{\arraystretch}{1.2}
  \begin{tabular*}{\hsize} { @{\extracolsep{\fill}} cccc} \hline \hline
    & $\Gamma_{\rm tot}$ (keV) & $\Gamma_{S=0}$ (keV) & $\Gamma_{S=1} $ (keV) \\ \hline
    full & ?88.2 & ?87.1 & ??1.1 \\
    $(l=odd)^2$ only & ?12.5 & ?10.7 & ??1.8 \\
    &&& \\
    no-pairing & 348.? & 232.? & 116.? \\
    ($ct=3000$ fm) &&& \\ \hline \hline
  \end{tabular*}
  \endgroup
  \catcode`? = 12 %initialize `?'.
  \caption{ The contributions from the spin-singlet and 
the spin-triplet configurations to the total decay width. 
Note that the experimental value of the total decay width is 
$92 \pm 6$ keV \cite{88Ajz,02Till}. 
All the values are evaluated at $ct=1200$ fm, 
except those in the ``no-pairing'' case, 
which are evaluated at $ct=3000$ fm. 
In all the cases, the total energy release 
(Q-value) of the two protons is set to be consistent to the 
experimental value, $1.37$ MeV. } \label{tb:2} \vspace{-3truemm}
\end{center} \end{table}
\begin{figure*}[t] \begin{center}
  \begin{tabular}{c} %switch-off the auto-turning
     \begin{minipage}{0.32\hsize} \begin{center}
     \fbox{ \includegraphics[height=38truemm, clip, trim = 0 0 0 0]
            {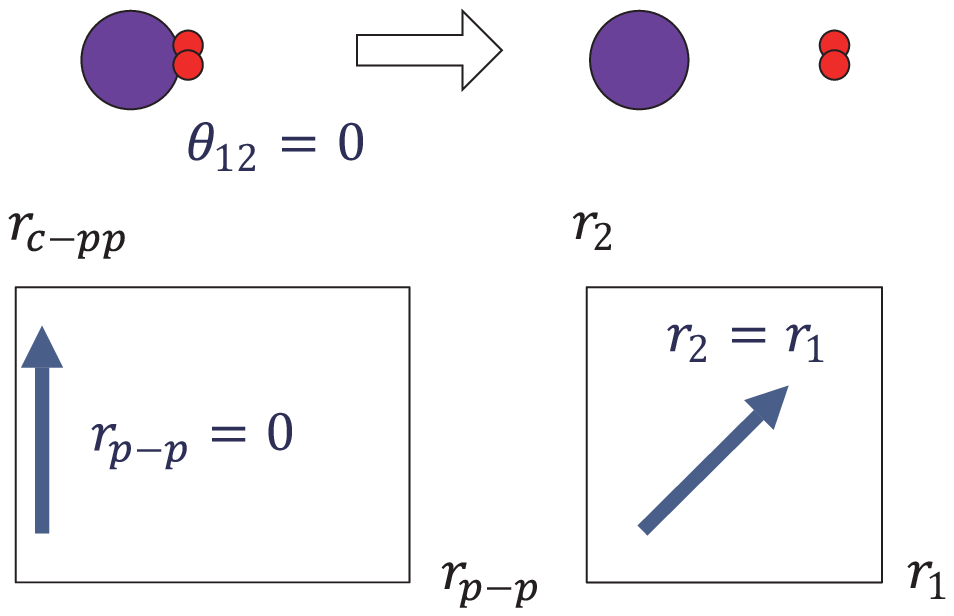}} \\ (a) diproton \\
     \vspace{3truemm}
     \fbox{ \includegraphics[height=38truemm, clip, trim = 0 0 0 0]
            {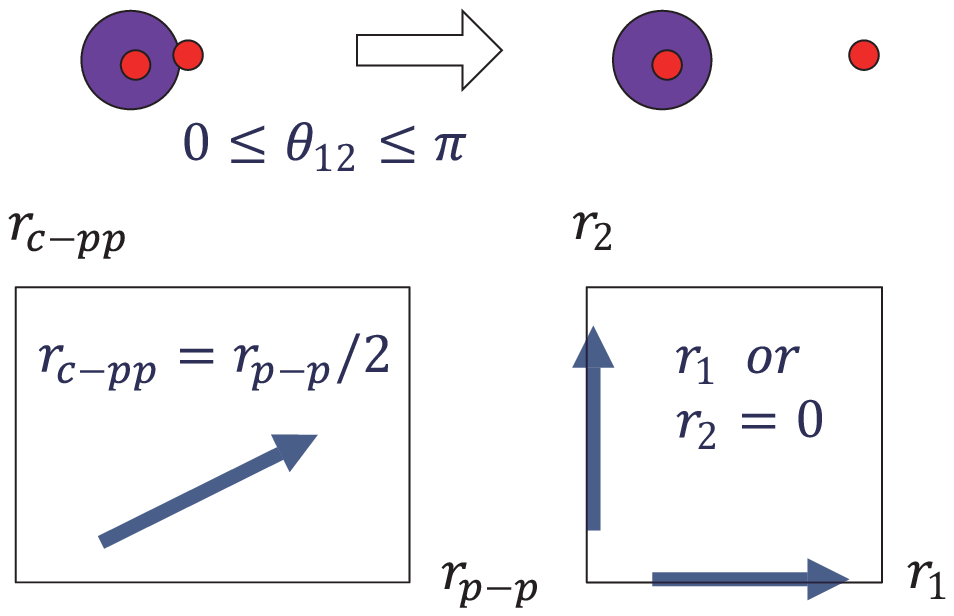}} \\ (d) one-proton \\
     \end{center} \end{minipage}

     \begin{minipage}{0.32\hsize} \begin{center}
     \fbox{ \includegraphics[height=38truemm, clip, trim = 0 0 0 0]
            {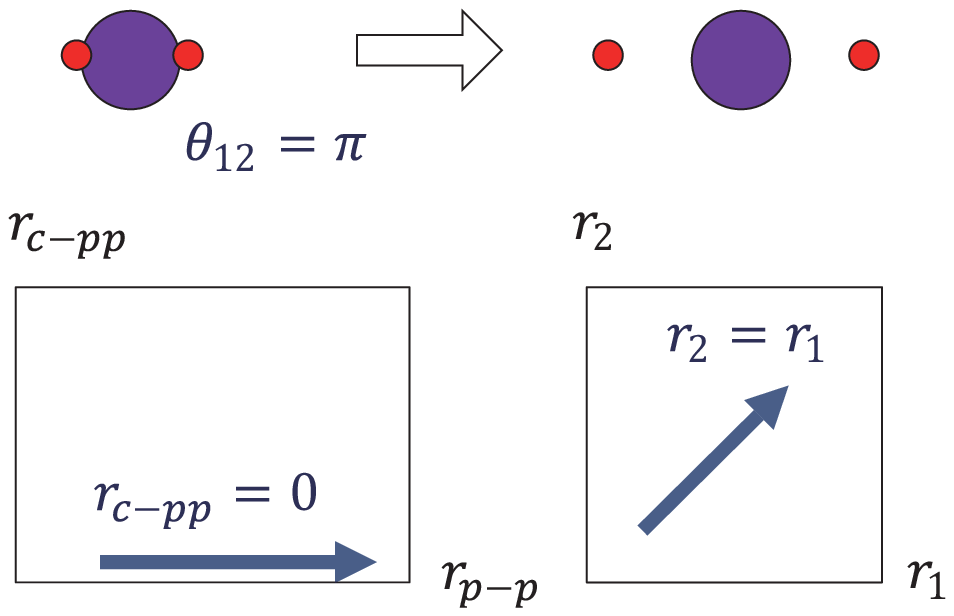}} \\ (b) simultaneous, $\theta_{12}=0$ \\
     \vspace{3truemm}
     \fbox{ \includegraphics[height=38truemm, clip, trim = 0 0 0 0]
            {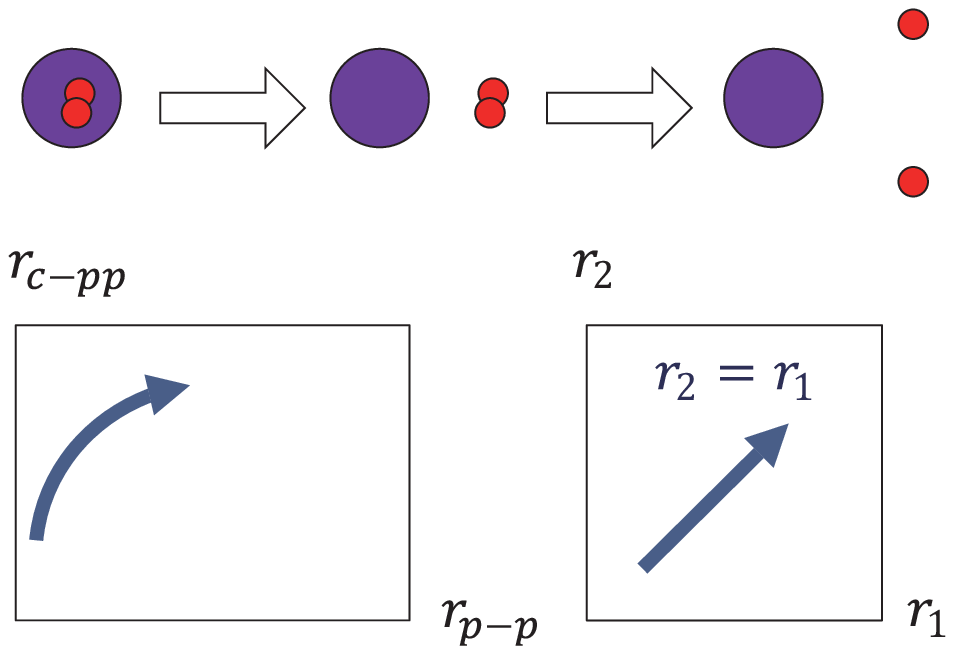}} \\ (e) correlated \\
     \end{center} \end{minipage}

     \begin{minipage}{0.32\hsize} \begin{center}
     \fbox{ \includegraphics[height=38truemm, clip, trim = 0 0 0 0]
            {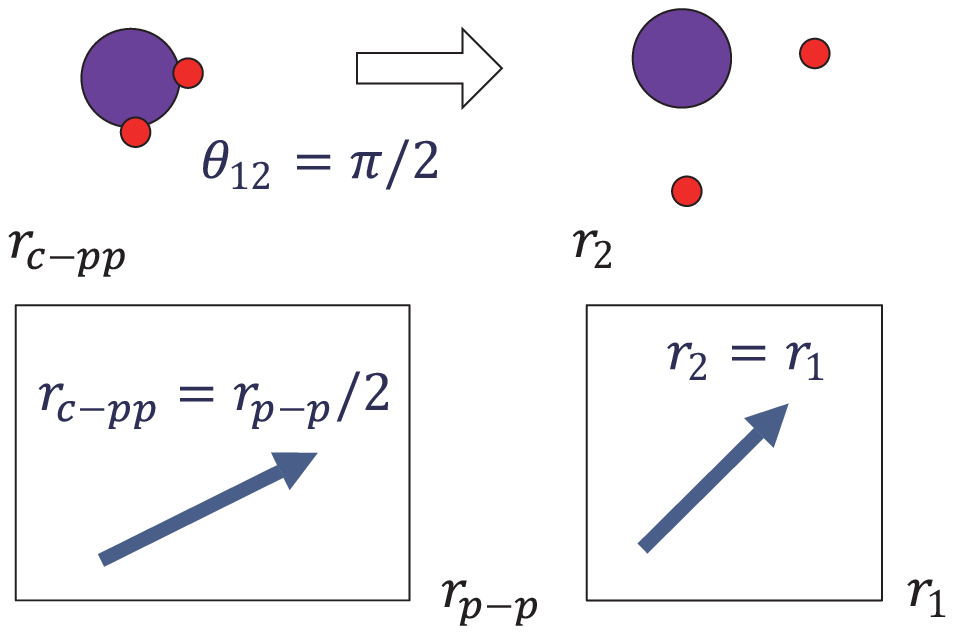}} \\ (c) simultaneous, $\theta_{12}=\pi/2$ \\
     \vspace{3truemm}
     \fbox{ \includegraphics[height=38truemm, clip, trim = 0 0 0 0]
            {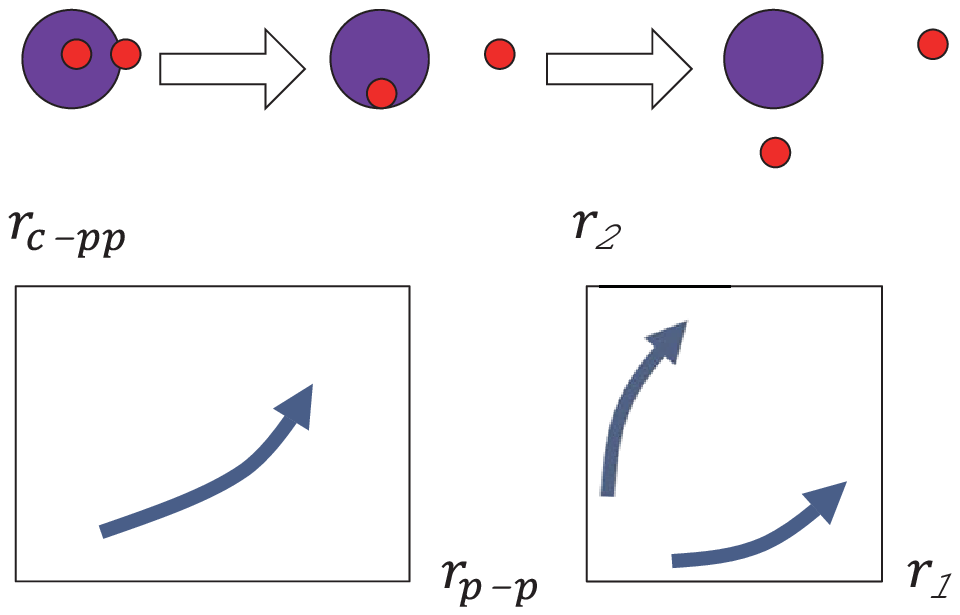}} \\ (f) sequential \\
     \end{center} \end{minipage}
  \end{tabular}
  \caption{(Color online) (a)-(c): Schematic illustrations for 
the trajectories of different \twop-emission modes. 
(d): The trajectory of the $1p$-emission. 
(e),(f): The same as the panels (a)-(d) but of the hybrid 
\twop-emissions. 
See the main text for the details. } \label{fig:80} \vspace{-3truemm}
\end{center} \end{figure*}

For the above two cases, 
we also calculate the partial decay widths for the spin-singlet and 
the spin-triplet configurations. 
The corresponding formula to Eq.(\ref{eq:pwidth}) is given as 
\beq
 \Gamma_{S}(t) \equiv \frac{\hbar }{1-N_d(t)} 
 \frac{d}{dt} N_{d,S}(t), 
\eeq
with 
\beqa
 && N_{d,S}(t) \equiv \Braket{\Psi_{d,S}(t) | \Psi_{d,S}(t)} \\
 && = \int_0^{R_{\rm box}} dr_1 \int_0^{R_{\rm box}} dr_2 
      \int_{0}^{\pi} d\theta_{12} \nonumber \\
 && \phantom{0000} \times 8\pi^2 r_1^2 r_2^2 \sin \theta_{12} 
    \abs{\Psi_{d,S} (t;r_1,r_2,\theta_{12})}^2. 
\eeqa
where $S$ indicates the combined spin of the two protons. 
The results are shown in Fig. \ref{fig:5-1}. 
Clearly, the spin-singlet configuration almost exhausts 
the decay width in the case of full configuration-mixture 
shown in the upper panel of Fig. \ref{fig:5-1}. 
This suggests that the emitted two protons from the ground state of $^6$Be 
have mostly the $S=0$ configuration like a diproton. 
On the other hand, in the lower panel of Fig. \ref{fig:5-1}, 
one can see that the spin-triplet configuration occupies a considerable 
amount of the total decay width when we exclude even-$l$ partial waves. 

In the first and the second rows of Table \ref{tb:2}, 
we tabulate the total and partial widths in 
the case of full configuration-mixture and in 
the $(l=odd)^2$ cases, respectively. 
The values are evaluated at $ct=1200$ fm, where the total widths 
sufficiently converge. 
Clearly, there is a significant increase of the spin-singlet width 
in the case of full configuration-mixture, 
by about one order of magnitude larger than that in 
the case of $(l=odd)^2$ waves. 
On the other hand, we get similar values of the spin-triplet width 
in these two cases. 
From this result, we can conclude that the mixture of 
the odd-$l$ and even-$l$ s.p. states is responsible 
for the enhancement of the spin-singlet emission, 
although the dominance of the spin-singlet configuration 
in the initial state is apparent in both the two cases. 

A qualitative reason for the dominance of the spin-singlet 
configuration is due to the $(s_{1/2})^2$ channel. 
Notice that the $(s_{1/2})^2$-channel is allowed only for $S=0$. 
%Considering the coupled orbit $L \equiv l_1 \oplus l_2$, 
%from the coupling rule to the spin-parity of $0^+$, 
%it is possible to take $S = L = 0$ from $l_1 = l_2 = 0$ 
%when $(s_{1/2})^2$ channel exists. 
Because there is no centrifugal barrier in this channel, 
the spin-singlet emission can be dominant. 
On the other hand, for the spin-triplet configuration, 
only $L = 1$ is permitted. 
Thus the $(s_{1/2})^2$ configuration does not contribute to it, and 
there is a centrifugal barrier for all the channels in 
the spin-triplet configuration. 
Consequently, apart from the reduction due to the stronger pairing attraction, 
the spin-triplet widths are similar in both the two cases. 

\subsection{Time-Evolution of Decay State}
\begin{figure*}[htb] \begin{center}
  \begin{tabular}{c} %switch-off the auto-turning
   \begin{minipage}{0.32\hsize} \begin{center}
     \fbox{ \includegraphics[height=35truemm, clip, trim = 15 0 15 25]{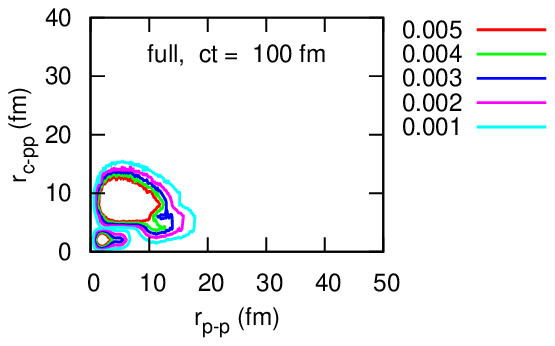}} \\
     \fbox{ \includegraphics[height=35truemm, clip, trim = 15 0 15 25]{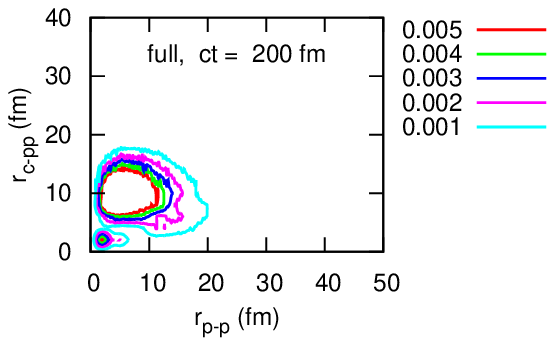}} \\
     \fbox{ \includegraphics[height=35truemm, clip, trim = 15 0 15 25]{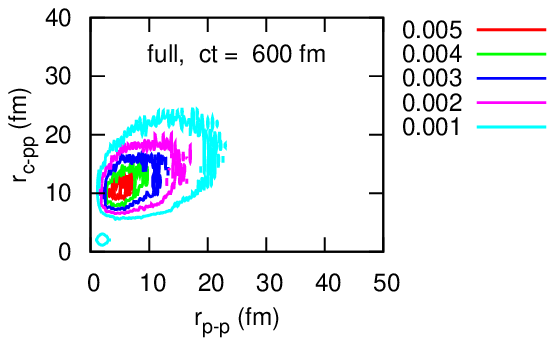}} \\
     \fbox{ \includegraphics[height=35truemm, clip, trim = 15 0 15 25]{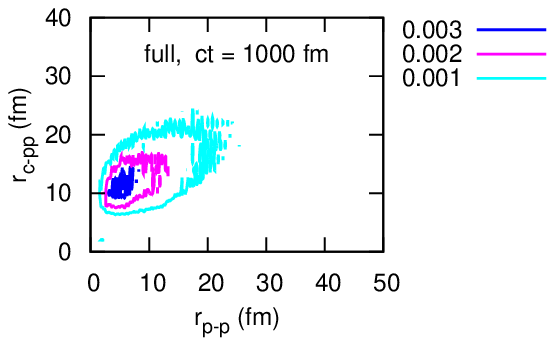}} \\
   \end{center} \end{minipage}

   \begin{minipage}{0.32\hsize} \begin{center}
     \fbox{ \includegraphics[height=35truemm, clip, trim = 20 0 10 25]{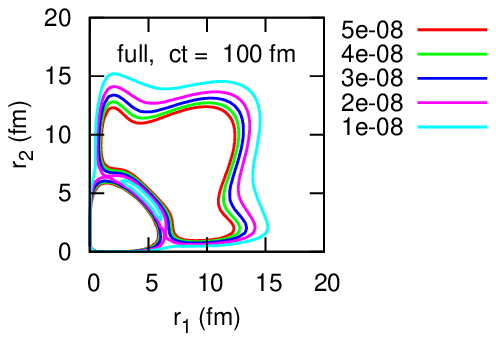}} \\
     \fbox{ \includegraphics[height=35truemm, clip, trim = 20 0 10 25]{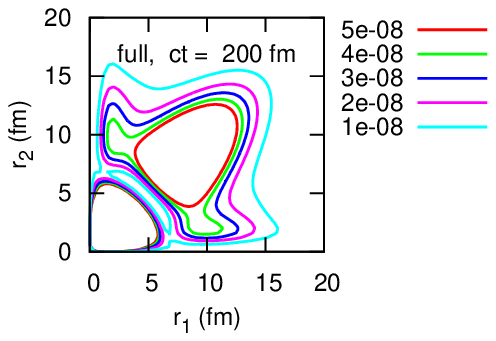}} \\
     \fbox{ \includegraphics[height=35truemm, clip, trim = 20 0 10 25]{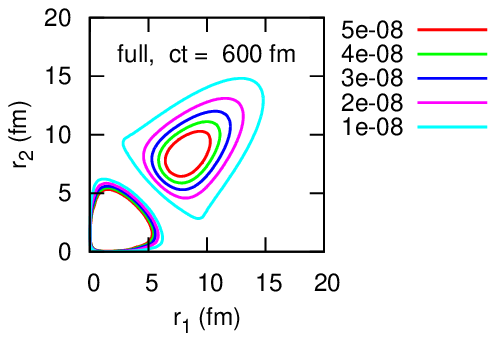}} \\
     \fbox{ \includegraphics[height=35truemm, clip, trim = 20 0 10 25]{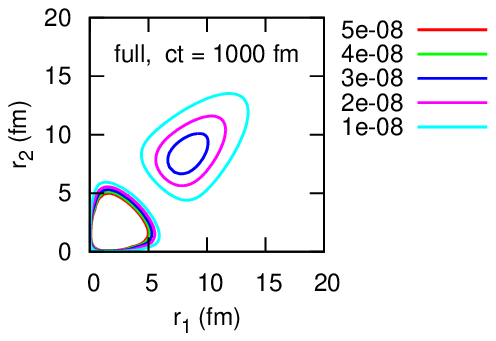}} \\
   \end{center} \end{minipage}

   \begin{minipage}{0.32\hsize} \begin{center}
     \fbox{ \includegraphics[height=35truemm, clip, trim = 10 5 0 5]{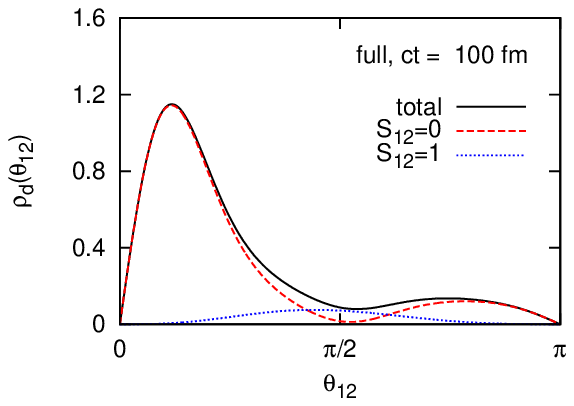}} \\
     \fbox{ \includegraphics[height=35truemm, clip, trim = 10 5 0 5]{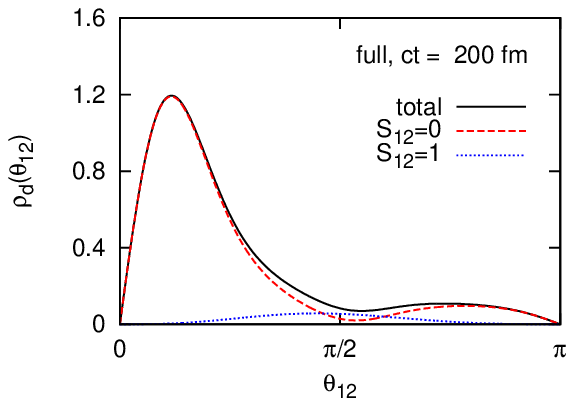}} \\
     \fbox{ \includegraphics[height=35truemm, clip, trim = 10 5 0 5]{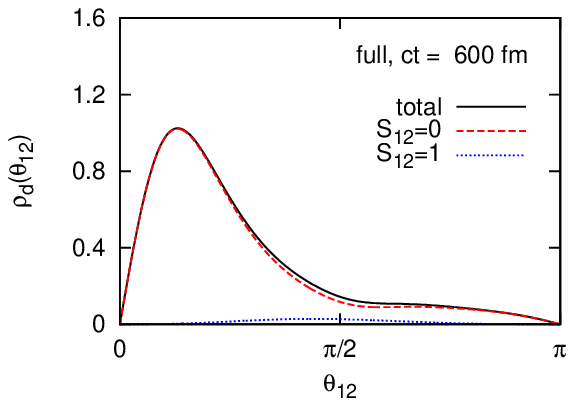}} \\
     \fbox{ \includegraphics[height=35truemm, clip, trim = 10 5 0 5]{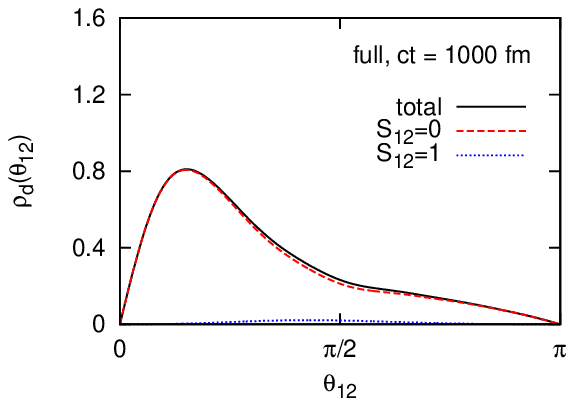}} \\
   \end{center} \end{minipage}
  \end{tabular}
  \caption{(Color online) 
The \twop-density distribution for the decay states, 
$\bar{\rho}_d(t)$, obtained with the time-dependent calculations. 
All the uncorrelated basis up to $(h_{11/2})^2$ are included. 
The left column: (i) the distribution as a function of 
$r_{\rm c-pp}$ %= (r_1^2 + r_2^2 + 2r_1r_2\cos \theta_{12})^{1/2}/2$ and 
and 
$r_{\rm p-p}$. % = (r_1^2+r_2^2-2r_1r_2\cos \theta_{12})^{1/2}$. 
The middle column: (ii) the distribution 
as a function of $r_1$ and $r_2$, 
obtained by integrating $\bar{\rho}_d$ for $\theta_{12}$. 
In order to clarify the peak(s), 
the radial weight $r_1^2 r_2^2$ is omitted. 
The right column: (iii) the angular distribution of the decay state 
plotted as a function of the opening angle $\theta_{12}$ 
between the two protons. 
It is obtained by integrating $\bar{\rho}_d(t)$ 
for the radial coordinates, $r_1$ and $r_2$. 
Beside the total distribution, the spin-singlet and spin-triplet components 
are also plotted. } \label{fig:81} \vspace{-3truemm}
\end{center} \end{figure*}

\begin{figure*}[htb] \begin{center}
  \begin{tabular}{c} %switch-off the auto-turning
   \begin{minipage}{0.32\hsize} \begin{center}
     \fbox{ \includegraphics[height=35truemm, clip, trim = 15 0 15 25]
            {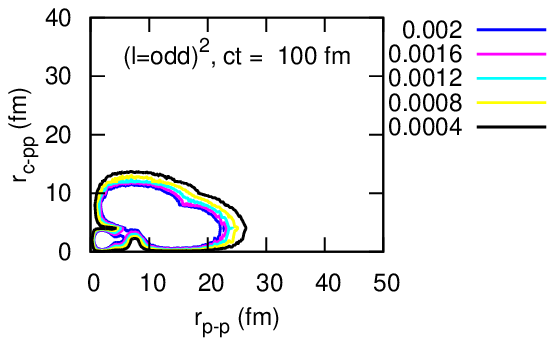}} \\
     \fbox{ \includegraphics[height=35truemm, clip, trim = 15 0 15 25]
            {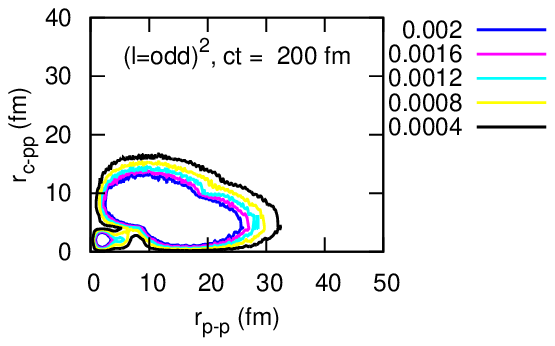}} \\
     \fbox{ \includegraphics[height=35truemm, clip, trim = 15 0 15 25]
            {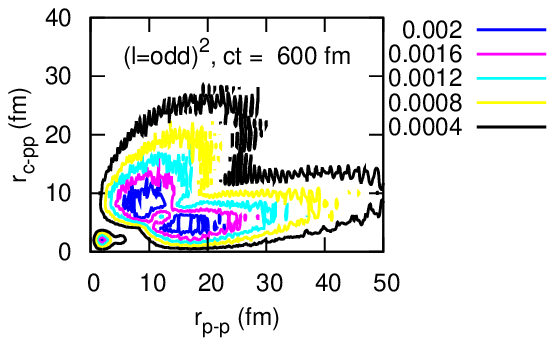}} \\
     \fbox{ \includegraphics[height=35truemm, clip, trim = 15 0 15 25]
            {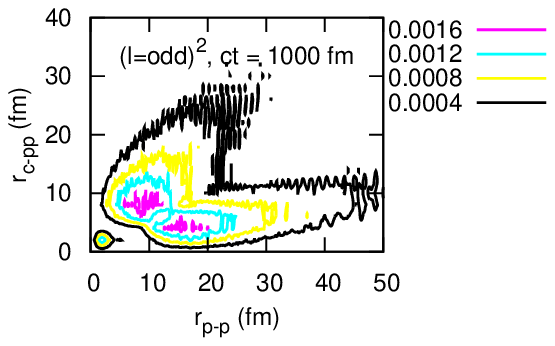}} \\
   \end{center} \end{minipage}

   \begin{minipage}{0.32\hsize} \begin{center}
     \fbox{ \includegraphics[height=35truemm, clip, trim = 20 0 10 25]
            {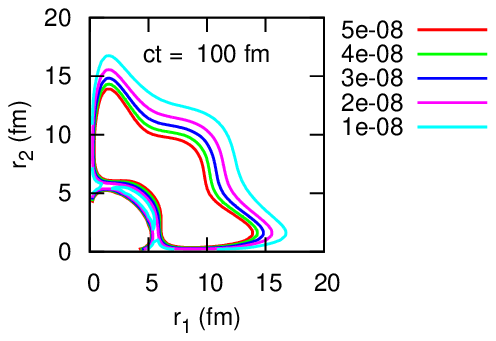}} \\
     \fbox{ \includegraphics[height=35truemm, clip, trim = 20 0 10 25]
            {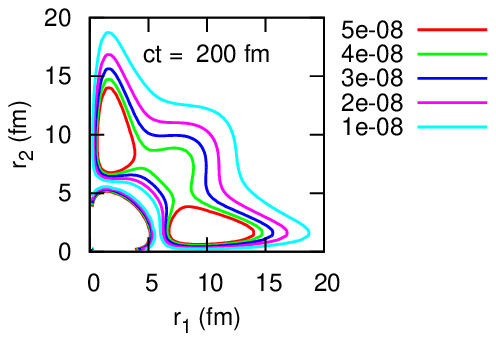}} \\
     \fbox{ \includegraphics[height=35truemm, clip, trim = 20 0 10 25]
            {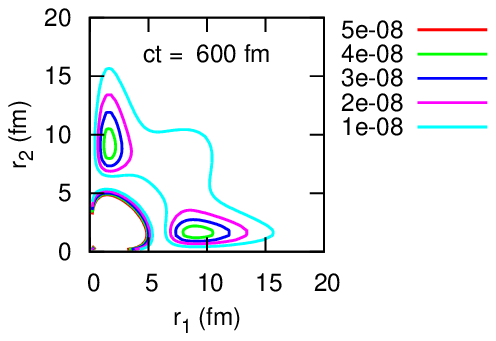}} \\
     \fbox{ \includegraphics[height=35truemm, clip, trim = 20 0 10 25]
            {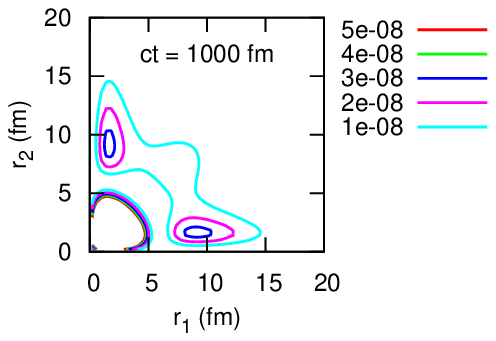}} \\
   \end{center} \end{minipage}

   \begin{minipage}{0.32\hsize} \begin{center}
     \fbox{ \includegraphics[height=35truemm, clip, trim = 10 5 0 5]
            {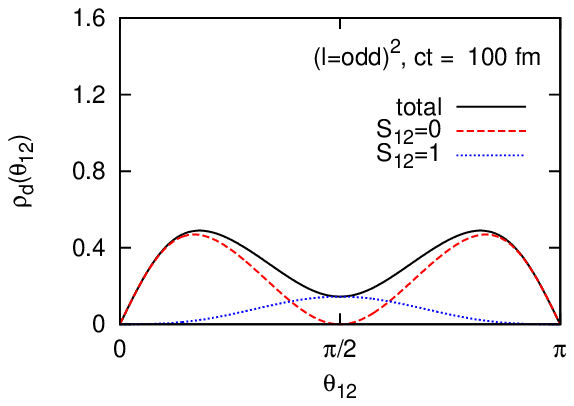}} \\
     \fbox{ \includegraphics[height=35truemm, clip, trim = 10 5 0 5]
            {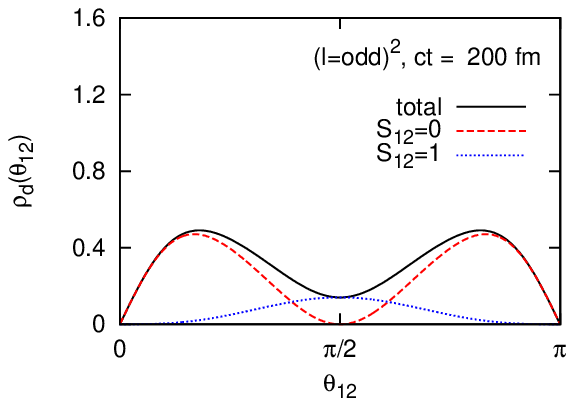}} \\
     \fbox{ \includegraphics[height=35truemm, clip, trim = 10 5 0 5]
            {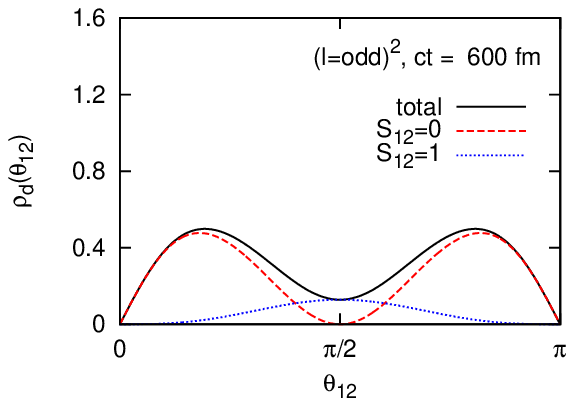}} \\
     \fbox{ \includegraphics[height=35truemm, clip, trim = 10 5 0 5]
            {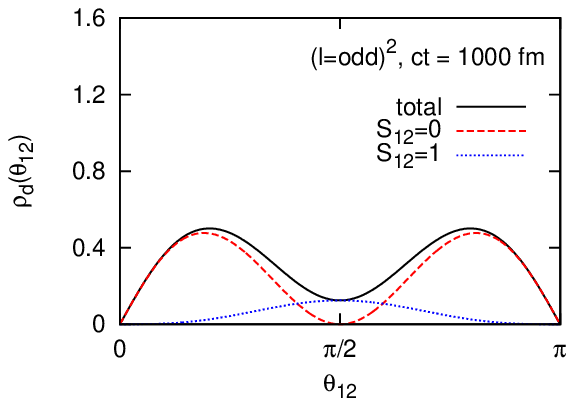}} \\
   \end{center} \end{minipage}
  \end{tabular}
  \caption{(Color online) 
The same as Fig.\ref{fig:81} but for the case with only $(l=odd)^2$ waves. 
Notice a different scale in the left column from that in 
Fig.\ref{fig:81}. } \label{fig:82} \vspace{-3truemm}
\end{center} \end{figure*}

In order to discuss the dynamics of the emission process, 
we show the density distribution of the decay state, 
\beqa
 && \bar{\rho}_{d}(t) = 8\pi^2 r_1^2 r_2^2 \sin \theta_{12} \rho_d(t), \\
 && \rho_d(t) = \abs{\Psi_{d} (t;r_1,r_2,\theta_{12})}^2. 
\eeqa
The decay state, which is orthogonal to the initial state 
confined inside the potential barrier, 
has the most of its amplitude outside the potential barrier. 
%The most of the amplitude of the decay state exists 
%outside the potential barrier, 
%because we prepare the initial state, which is orthogonal to the decay state, 
%so as to have no amplitude in that region. 
%For the presentation, we renormalize the $\bar{\rho}_d(t)$ so that 
%its integration become unity at each time: 
%\beq
%  \bar{\rho}_d (t) \longrightarrow \frac{\bar{\rho}_d (t)}{N_d (t)}, 
%\eeq
%where $N_d (t)$ is the decay probability given by Eq.(\ref{eq:603DComp}). 
In the following, 
we adopt three sets of radial coordinates: 
(i) The first set includes 
$r_{\rm c-pp}$ %= (r_1^2 + r_2^2 + 2r_1r_2\cos \theta_{12})^{1/2}/2$ 
and $r_{\rm p-p}$, % = (r_1^2+r_2^2-2r_1r_2\cos \theta_{12})^{1/2}$, 
similarly to the left panel of Fig. \ref{fig:2}. 
(ii) In the second set, we integrate $\bar{\rho}_d$ with respect to 
the opening angle, $\theta_{12}$, 
and plot it as a function of $r_1$ and $r_2$. 
In order to see the peak-structure clearly, 
we omit the radial weight $r_1^2 r_2^2$ in 
$\bar{\rho}_d$ in the second set. 
(iii) Within the third set, on the other hand, 
we integrate $\bar{\rho}_d(t)$ over the radial distances, 
and plot it as a function of $\theta_{12}$. 
We will use in Figs. \ref{fig:81}, \ref{fig:82} and \ref{fig:83} 
these sets of coordinates in order to present the amplitude 
of the decay state in actual calculations. 

Before we show the results of the actual calculations, 
we schematically illustrate the dynamic of the 
\twop-emissions in Fig. \ref{fig:80}. 
From the geometry, the emission modes are classified into two categories: 
``simultaneous two-proton'' and ``one-proton'' emissions. 
The diproton emission is a special case of the first category. 
The second category corresponds to the case 
where only one proton penetrates the barrier. 
%The trajectories of three simultaneous \twop- and a $1p$-emissions 
%are schematically shown in Fig. \ref{fig:80}(a), (b), (c) and (d). 

Figs. \ref{fig:80}(a), (b) and (c) correspond to the simultaneous 
\twop-emissions with $\theta_{12}=0,\pi$ 
and $\pi/2$, respectively, where $\theta_{12}=0$ (Fig.\ref{fig:80}(a)) 
corresponds to the diproton emission. 
In these three cases, the density in the $(r_1,r_2)$-plane shows the same 
patterns, and is concentrated along $r_1\cong r_2$. 
The simultaneous emissions with different opening angles 
 can be distinguished only in the $(r_{\rm p-p}, r_{\rm c-pp})$-plane: 
for instance, in the diproton emission, 
the probability shows mainly along the line with 
$r_{\rm c-pp} \gg r_{\rm p-p}$, 
while it is along the line with $r_{\rm c-pp}=0$ for $\theta_{12}=\pi$. 
In the one-proton emission shown in Fig.\ref{fig:80}(d), 
only one of the two protons goes through the barrier while 
the other proton remains inside the core nucleus. 
This is seen as the increment along $r_{\rm c-pp} \cong r_{\rm p-p}/2$ 
and $r_1$ or $r_2 \cong 0$ lines. 

In Fig. \ref{fig:80}(e) and (f), 
we illustrate two hybrid processes. 
The first one is a ``correlated emission'', 
shown in Fig. \ref{fig:80}(e). 
In the correlated emission, 
the two protons are emitted simultaneously 
to almost the same direction, 
holding the diproton-like configuration. 
In this mode, in the early stage of tunneling, 
the density distribution 
has a larger amplitude in the region with 
$r_1 \cong r_2$ and small $\theta_{12}$. 
In the $(r_{\rm p-p}, r_{\rm c-pp})$-plane, 
it corresponds to the increment of the probability 
in the region of $r_{\rm p-p} \ll r_{\rm c-pp}$. 
After the barrier penetration, the 
two protons separate from each other 
mainly due to the Coulomb repulsion. 
The second hybrid process is a ``sequential emission'', 
which is shown in Fig. \ref{fig:80}(f). 
In this mode, there is a large possibility in which one proton is emitted 
whereas the other proton remains around the core. 
The density distribution shows high peaks along 
$r_1 \gg r_2$ and $r_1 \ll r_2$. 
In the $(r_{\rm p-p}, r_{\rm c-pp})$-plane, it corresponds to 
the increment along the line of 
$r_{\rm c-pp} \cong r_{\rm p-p}/2$. 
In contrast to 
the pure one-proton emission, 
the remaining proton eventually goes 
through the barrier 
when the core-proton subsystem is unbound. 

\subsubsection{case of full configuration-mixture}
We now show the results of the time-dependent calculations 
for the \twop-emission of $^6$Be. 
We first discuss the case of full configuration-mixture, 
where the odd-$l$ and even-$l$ single particle states are fully mixed 
by the pairing correlation. 
The density distribution for the decay state 
along the time-evolution is shown in Fig. \ref{fig:81}. 
The left, middle and right columns correspond to the coordinate sets 
(i), (ii) and (iii) defined before, respectively. 
The first to the fourth panels in each column show the decay-density at 
$ct=100,200,600$ and $1000$ fm, respectively. 
For a presentation purpose, we normalize $\bar{\rho}_d$ 
at each step of time. 

In the left and middle columns of Fig. \ref{fig:81}, 
it can be seen that the process in this case 
is likely the correlated emission shown in Fig. \ref{fig:80}(e). 
Contributions from the other modes shown 
in Fig. \ref{fig:80} are small. 
In the middle column of  Fig. \ref{fig:81}, 
during the time-evolution, 
there is a significant increment of 
$\bar{\rho}_d$ along the line with $r_1 \cong r_2$. 
The corresponding peak in the left column is at 
$r_{\rm p-p} \ll r_{\rm c-pp} \cong 10$ fm, which means 
a small value of $\theta_{12}$. 
It should also be noted that, after the barrier penetration, 
the two protons lose their diproton-like configuration 
due to the Coulomb repulsion, which results in the increase of $r_{\rm p-p}$. 
Thus, for $r_{\rm c-pp} \geq 10$ fm which is 
a typical position of the potential barrier from the core, 
the density distribution extends around 
the $r_{\rm c-pp} \cong r_{\rm p-p}$ region. 
In this process, the pairing correlation plays an important role 
to generate the significant diproton-like configuration before 
the end of the barrier penetration. 
In the right column of Fig. \ref{fig:81}, the distributions 
are also displayed as a function of 
the opening angle, $\theta_{12}$. 
We can clearly see that the decay state has a high peak at 
$\theta_{12} \cong \pi/6$ in this time-region. 
%The emitted two protons should show the opening angles close to this value. 

These results imply that the two protons are emitted almost in the same 
direction, at least in the early stage of the emission process. 
Intuitively, from the uncertainty principle, this would correspond to 
a large opening angle in the momentum space. 
Indeed, such component has been experimentally observed to be 
dominant for the \twop-decay of $^{6}$Be \cite{09Gri_80, 09Gri_677}. 
It would be an interesting future work to carry out the Fourier 
transformation of the decay state and compare our calculations with 
the experimental data. 

\subsubsection{$(l=odd)^2$ case}
We next discuss the case only with 
$(l=odd)^2$ bases. 
In Fig. \ref{fig:82}, 
the decay density shows a strong pattern of the sequential emission 
demonstrated in Fig. \ref{fig:80}(f): 
significant increments occur along the lines with 
$r_{\rm c-pp} \cong r_{\rm p-p}/2$ and $r_1 \gg r_2$ or $r_1 \ll r_2$. 
Notice that the contribution from the simultaneous 
emissions also exists, especially in the early time-region. 
As a result, the decay state has widely spread amplitudes 
as a mixture of these emission modes. 
However, the simultaneous mode is minor compared with 
the case of full configuration-mixture. 
Notice that the condition for a true \twop-emitter is satisfied also 
in this case: 
the core-proton resonance is located at $1.96$ MeV 
which is above $Q_{\rm 2p}=1.37$ MeV. 
However, even with the strong pairing attraction and the energy 
condition for the true \twop-emitter, 
the process hardly becomes the correlated emission when the 
parity-mixing is forbidden or extensively suppressed. 
The angular distribution shows exactly the symmetric form, 
and is almost invariant during the time-evolution. 
This is because we exclude the pairing correlation 
between the positive and negative parity states in the core-proton system, 
not only at $t=0$ but also during the time-evolution. 
%In other words, there are almost no FSIs in this case, which 
%alter the shape of the angular distribution. 

\subsubsection{no-pairing case}
\begin{figure}[b] \begin{center}
  \fbox{ \includegraphics[width=0.8\hsize, clip, trim = 10 0 0 5]{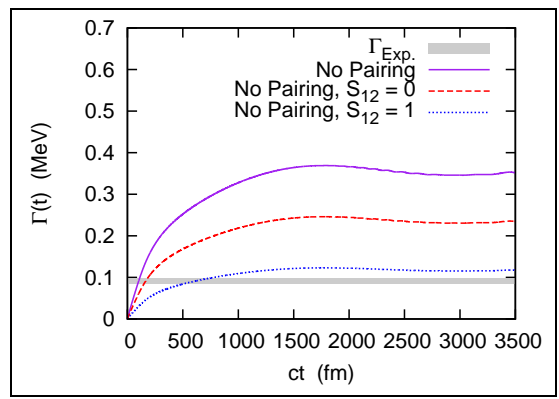}}
  \caption{(Color online) The same as Fig. \ref{fig:5-1} but for the case 
without pairing correlations. } \label{fig:5-3} \vspace{-3truemm}
\end{center} \end{figure}

\begin{figure*}[t] \begin{center}
  \begin{tabular}{c} %switch-off the auto-turning
   \begin{minipage}{0.32\hsize} \begin{center}
     \fbox{ \includegraphics[height=35truemm, clip, trim = 15 0 15 25]
            {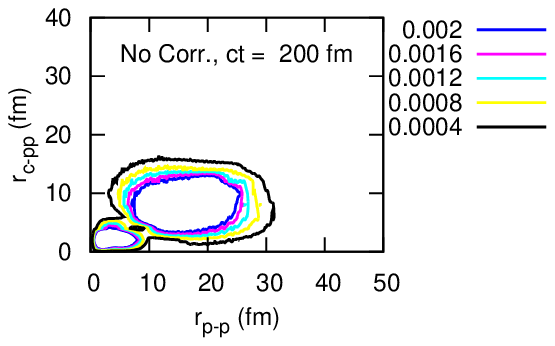}} \\
     \fbox{ \includegraphics[height=35truemm, clip, trim = 15 0 15 25]
            {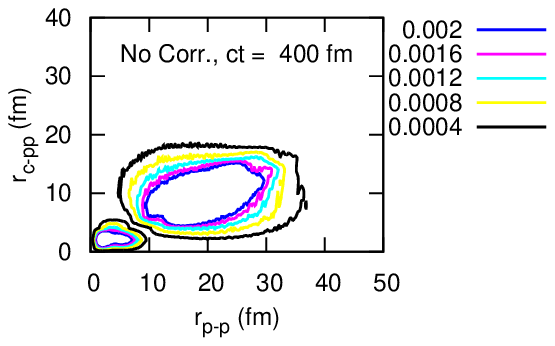}} \\
     \fbox{ \includegraphics[height=35truemm, clip, trim = 15 0 15 25]
            {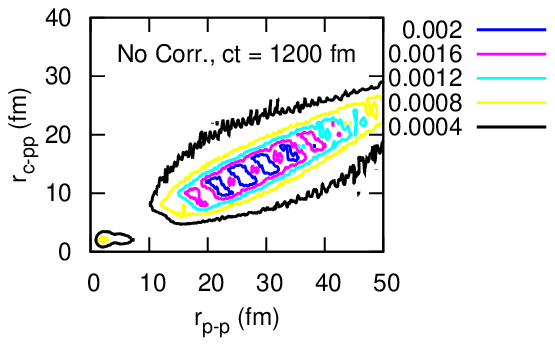}} \\
     \fbox{ \includegraphics[height=35truemm, clip, trim = 15 0 15 25]
            {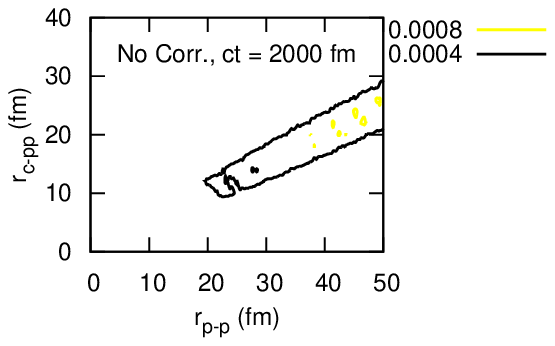}} \\
   \end{center} \end{minipage}

   \begin{minipage}{0.32\hsize} \begin{center}
     \fbox{ \includegraphics[height=35truemm, clip, trim = 20 0 10 25]
            {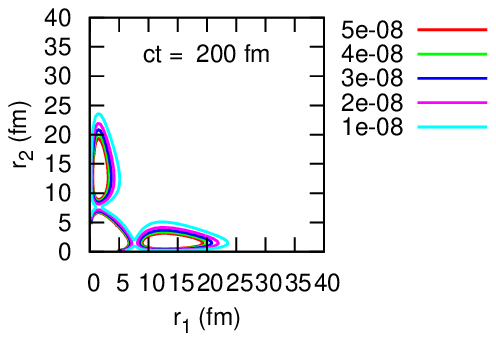}} \\
     \fbox{ \includegraphics[height=35truemm, clip, trim = 20 0 10 25]
            {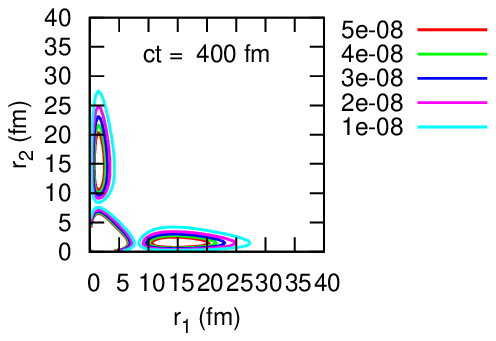}} \\
     \fbox{ \includegraphics[height=35truemm, clip, trim = 20 0 10 25]
            {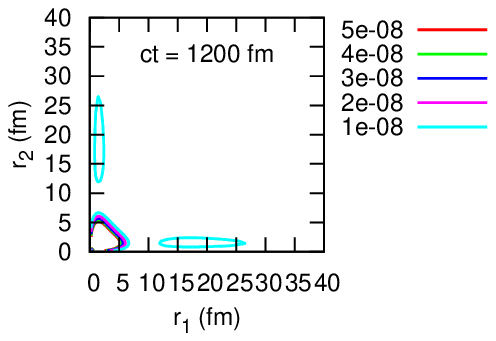}} \\
     \fbox{ \includegraphics[height=35truemm, clip, trim = 20 0 10 25]
            {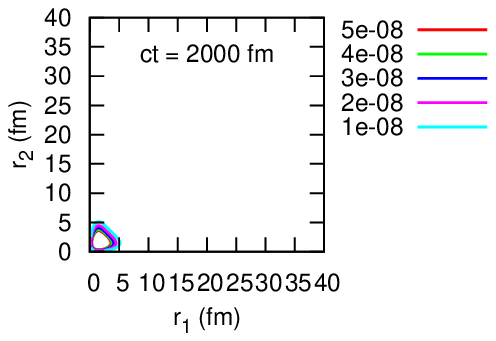}} \\
   \end{center} \end{minipage}

   \begin{minipage}{0.32\hsize} \begin{center}
     \fbox{ \includegraphics[height=35truemm, clip, trim = 10 5 0 5]
            {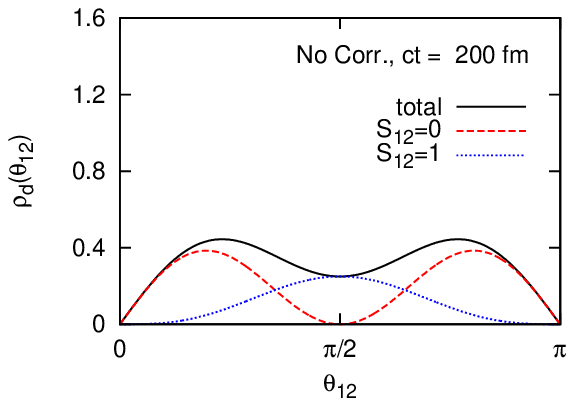}} \\
     \fbox{ \includegraphics[height=35truemm, clip, trim = 10 5 0 5]
            {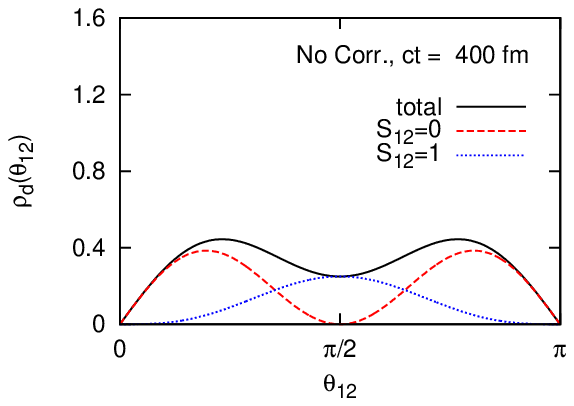}} \\
     \fbox{ \includegraphics[height=35truemm, clip, trim = 10 5 0 5]
            {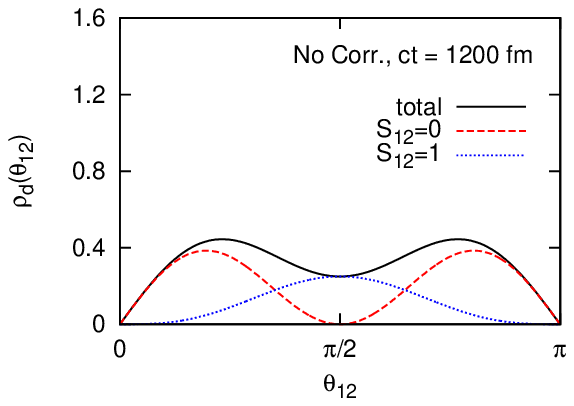}} \\
     \fbox{ \includegraphics[height=35truemm, clip, trim = 10 5 0 5]
            {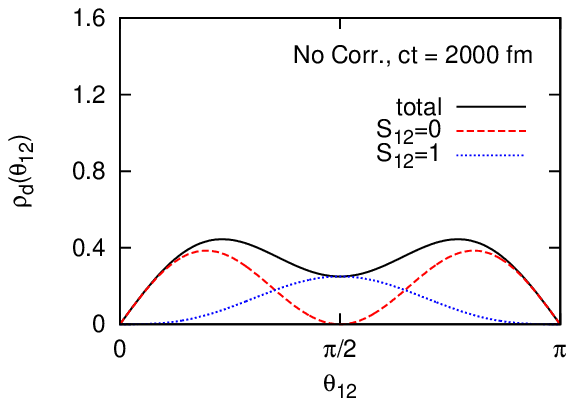}} \\
   \end{center} \end{minipage}
  \end{tabular}
  \caption{(Color online) The same as Fig. \ref{fig:81} but for the case 
without pairing correlations. } \label{fig:83} \vspace{-3truemm}
\end{center} \end{figure*}

Finally, for a comparison with the above two cases, 
we also perform similar calculations but by 
completely neglecting the pairing correlation. 
In this case, we only consider the uncorrelated 
Hamiltonian, $h_1 + h_2$. 
Because of the absence of the non-diagonal components 
in the Hamiltonian matrix, 
it can be proved that, if the s.p. resonance is at 
an energy $\epsilon_0$ with its width $\gamma_0$, 
the \twop-resonance is at $2 \epsilon_0$ with its 
width $2 \gamma_0$. 
The \twop-wave function is expanded on the uncorrelated 
basis with a single set of angular quantum numbers. 
Namely, 
\beq
 \ket{\Psi_{(lj)}(t)} 
 = \sum_{n_a,n_b} C_{n_a n_b}(t) \ket{{\Phi}_{n_a n_b lj}}, 
\eeq
where $(lj)=p_{3/2}$ for $^6$Be. 
In order to reproduce the empirical Q-value of $^6$Be, 
we inevitably modify the core-proton potential. 
We employ $V_0=-68.65$ MeV 
to yield the s.p. resonance at $\epsilon_0(p_{3/2}) = 1.37/2 = 0.685$ MeV, 
although the scattering data for the core-proton subsystem 
are not reproduced and 
the character of a true \twop-emitter disappears. 
With this potential, we obtain the s.p. resonance with 
a broad width: $\gamma_0(p_{3/2}) \cong 170$ keV. 
Because of the broad decay width, 
we need to increase the radial box to $R_{\rm box}=200$ fm 
in order to neglect the artifact due to the reflection at 
$R_{\rm box}$ in the long time-evolution. 

The result for the decay width is shown in Fig. \ref{fig:5-3} and in 
the last row of Table \ref{tb:2}. 
To get the saturated result, 
we somewhat need a relatively longer time-evolution than that 
in the case of full configuration-mixture. 
Thus, in Table \ref{tb:2}, we evaluate the decay width 
at $ct=3000$ fm. 
By this time, the total decay width, $\Gamma(t)$, converges to about $340$ keV 
which is consistent to that expected from 
the s.p. resonance, $\gamma_0(p_{3/2})$. 
During the time-interval shown in Fig. \ref{fig:5-3}, 
there still remain some oscillations in $\Gamma(t)$. 
This is a characteristic behavior of the broad resonance, 
namely an oscillatory deviation from the exponential decay-rule. 
For the spin-singlet and triplet configurations, 
their contributions have exactly the ratio of $2:1$. 
This result is simply due to 
the re-coupling of the angular momentum for 
the $(p_{3/2})^2$ configuration. 
%We also expand the radial box to $R_{\rm box}=200$ fm in order 
%to neglect the artifact due to the reflection 
%in the longer time-evolution. 

By comparing these results with those in 
the case of the full configuration-mixing, 
we can clearly see a decisive role of the pairing correlations 
in \twop-emissions. 
Assuming the empirical Q-value, 
if we explicitly consider the pairing correlations, 
the decay width becomes narrow and agrees with 
the experimental data. 
On the other hand, in the no-pairing case, 
we need a modified core-proton interaction to reproduce the 
empirical Q-value, and the properties of 
the core-proton resonance state become inconsistent 
with the experimental data. 
Even though the Q-value is adjusted in this way, 
the calculated \twop-decay width 
is significantly overestimated in this case. 
Namely, we cannot simultaneously reproduce 
the experimental Q-value and the decay width with 
the no-pairing assumption. 
If one is forced to reproduce them simultaneously, 
one may need an unphysical 
core-proton interactions. 

In Fig. \ref{fig:83}, we show the density distribution 
of the decay state during the time-evolution. 
Obviously, in this case, 
the process is the sequential or, moreover, like 
the one-proton emission. 
There is a significant increase of the density 
along the lines with $r_{\rm c-pp} \cong r_{\rm p-p}/2$ and, consistently, 
with $r_1 \gg r_2$ and $r_1 \ll r_2$ (see Fig. \ref{fig:80}). 
On the other hand, the probability for the simultaneous and 
correlated emissions are negligibly small. 
This is quite different from that in the 
case of full configuration-mixture, 
where the correlated emission is apparent. 
%Notice that, with a disagreement with the experimental decay width, 
%this result should not correspond to 
%the \twop-emission of $^6$Be in reality. 
%This situation can be interpreted as the limit where the 
%core-proton resonance plays an excessively dominant role. 

\subsection{Role of Pairing Correlation in Decay Width}
In this subsection, we discuss a general role of the pairing correlation 
in the \twop-emission. 
%% HS %%
To this end, we calculate the \twop-decay width for different Q-values 
in the case of full configuration-mixture and the no-pairing case. 
The variation of the Q-value is done by changing the parameter $V_0$ in 
the core-proton potential (Eq.(\ref{eq:cp-WS})), 
while the pairing interaction used in the case of 
full configuration mixture is kept unchanged. 
Notice that for the no-pairing case, the s.p. resonance 
appears at $\epsilon_0(p_{3/2}) =Q/2$. 
%% HS %%
%% TO %%
%The variation of the Q-value is done by changing the parameter $V_0$ in 
%the core-proton potential (Eq.(\ref{eq:cp-WS})). 
%In the previous calculations, we used $V_0=-58.7$ and $V_0=-68.65$ MeV 
%in the case of full configuration mixture and 
%the no-pairing cases, respectively, 
%in order to reproduce the empirical Q-value, $Q_{\rm 2p}=1.37$ MeV. 
%At this moment, in the no-pairing case, 
%we also adopt four points for plotting, indicated by 
%$V_0=-68.65 \pm 0.5$ and $V_0=-68.65 \pm 1.0$ MeV. 
%In the case of full configuration mixture, on the other hand, 
%we additionally adopt four points indicated by 
%$V_0=-58.7 \pm 0.5$ and $V_0=-58.7 \pm 1.0$ MeV, 
%while the pairing interaction is kept unchanged. 
%% TO %%
\begin{figure}[tb] \begin{center}
  \begin{tabular}{c} %switch-off the auto-turning
  \fbox{ \includegraphics[width=0.98\hsize, clip, trim = 10 5 0 5]
         {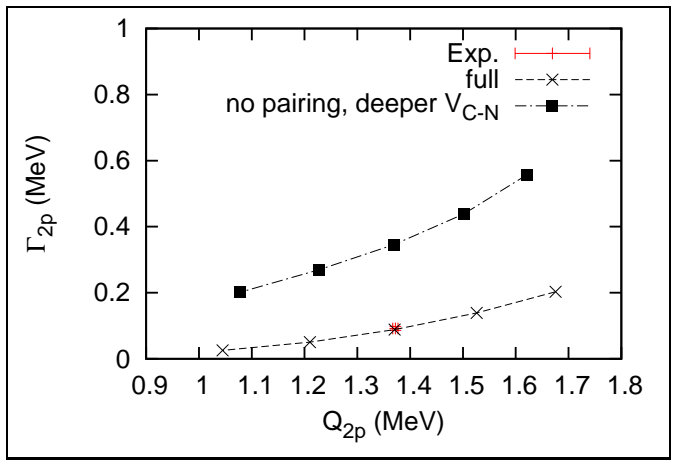}}
  \end{tabular}
  \caption{The calculated decay width for the 
\twop-emission of $^{6}$Be, as a function of the Q-value. 
The Q-value is varied by modifying the core-proton potential. 
The experimental values, $Q_{\rm 2p}=1.372(5)$ MeV and 
$\Gamma_{\rm 2p}=0.092(6)$ MeV \cite{NNDCHP}, 
are also indicated. } \label{fig:5qp}
\end{center} \end{figure}

In Fig. \ref{fig:5qp}, 
the decay width is plotted as a function of the decay Q-value. 
We note that the calculated decay widths are well converged after 
a sufficient time-evolution in all the cases. 
The decay width is evaluated at 
$ct=1200$ and $3000$ fm in the full-correlation and 
the no-pairing cases, respectively. 
Clearly, the no-pairing calculations overestimate the decay width, 
in all the region of $Q_{\rm 2p}$. 
Namely, the three-body system becomes easier to decay without 
the pairing correlation, 
for the same value of the total energy release (Q-value). 
In other words, the pairing correlation plays an essential role in the 
meta-stable state, stabilizing it against particle emissions. 
We note that a similar effect has been predicted also for 
a one-neutron resonance, that is, 
the width of a one-neutron resonance becomes narrow 
when one considers the pairing correlations \cite{14Kobayashi}. 

Also, as we have confirmed in the previous section, 
the emission dynamics with and without the 
pairing correlations are essentially different to each other: 
the correlated emission becomes dominant if the pairing correlation 
is fully considered, 
whereas the sequential emission plays a major role in the no-pairing case. 
%Of course, this result can be associated with the character of $^6$Be 
%as a true \twop-emitter. 
Consequently, the pairing correlation must be 
treated explicitly in the meta-stable states, 
otherwise one would miss the essential effect on both the 
decay-rule and the dynamical phenomena. 

%%%%%%%%%%%%%%%%%%%%%%%%%%%%%%%%%%%%%%%%%%%%%%%%%%%%%%%%%%%%%%%%%%%%%%%%%%%
\section{Summary} \label{Sec:sum}
We have investigated the \twop-emission of the $^6$Be nucleus 
by employing a three-body model consisting of an $\alpha$ particle 
and two valence protons. 
We have 
applied the time-dependent method 
and discussed the decay dynamics of many-body 
meta-stale states, particularly in connection to the diproton 
correlation. 
An advantage of 
the time-dependent method is that 
it provides not only a way to evaluate the decay width, but also 
an intuitive way to understand the decay dynamics. 

By using the confining potential method, 
we first obtained the initial state of $^6$Be, in which 
the two protons are confined inside the potential barrier. 
Because of the pairing correlation between the two protons, 
the initial configuration includes the diproton correlation, 
similarly to the dineutron correlation in the ground states of 
Borromean nuclei such as $^6$He. 
At time $t=0$, the confinement is 
removed so that the \twop-state evolves in space and time. 
In this calculation, the decay width can be read off by 
plotting the survival probability as a function of time. 
We have found that our Hamiltonian well reproduces simultaneously 
the experimental Q-value and decay width of $^6$Be. 
We have also shown that the decay state predominantly 
has the spin-singlet configuration. 

By monitoring the time-evolution of the 
density distribution of the decay state, 
we have confirmed that 
the decay process in the early stage is mainly the correlated emission, 
in which the two protons tend to be 
emitted in a similar direction, 
reflecting the diproton correlation in the initial state. 
Thus, the \twop-emission can be a promising tool to probe experimentally the 
diproton correlation. 
We have also performed the calculations 
by including only odd-$l$ partial waves in order to switch 
off the diproton correlation. 
In this case, even though we use the model parameters which 
reproduce the empirical Q-value, 
the decay width is significantly underestimated. 
The decay process shows a large component of the sequential emission, 
in contrast to the case of full configuration-mixture. 
From these results, we can conclude that the diproton correlation 
plays an important role in the \twop-emission, providing an opportunity 
to probe it by observing the \twop-emission. 

We have also checked that, if the pairing correlation is completely neglected, 
the decay width is 
largely overestimated, partly because 
the proton-core potential has to be made deeper in order to yield 
the empirical Q-value. 
By monitoring the time-evolution of the 
density distribution of the decay state, 
it has been clarified that the emission is mostly a sequential decay 
with the no-pairing assumption. 
Namely, the pairing correlation is critically important to determine 
not only the decay width but also the dynamical phenomena.

In order to compare quantitatively the calculated results with 
the experimental data, we would need a more careful treatment of 
the final-state interactions (FSIs). 
In this work, we mainly treat the early stage of the time-evolution, 
terminating the calculations at $ct\sim 1400$ fm in order to avoid 
the artifact due to the reflection at the edge of the box, $R_{\rm box}$. 
On the other hand, the two protons are detected in the actual experiments at 
a much later time after being significantly affected by the FSIs. 
In order to fully take into account the FSIs, we would 
have to use an extremely large box even though the 
computational costs would increase severely. 

The time-dependent method which we employed in this paper 
can be applied also to a decay of other 
many-body meta-stable states. 
It provides a novel and intuitive point of view to 
the decay process. 
It would be an interesting future problem 
to apply this method to other problems of many-particle quantum 
decays, such as the two-neutron emission and the two-electron 
auto-ionization of atoms. 

%Additionally, in future works, 
%it will be necessary to concern the pairing interaction. 
%The pairing interaction employed in this paper 
%should be regarded as an effective interaction, 
%since it is inconsistent to the scattering 
%problem of \twop in vacuum due to our modification of $v_0$. 
%Within further expanded model space, it may cause the unphysical result. 
%One may also introduce a three-body force, 
%which works only if three particles 
%are close to each other \cite{01Myo,10Kiku,95Aoyama}. 
%The effect of this three-body force on decay processes is an 
%important topic, in regard to whether such an interaction is 
%really just a phenomenological one 
%or has a physical meaning beyond the two-body force. 
%Work towards this direction is in progress now. 

\begin{acknowledgments}
We thank M. Matsuo and R. Kobayashi for useful discussions on the 
effect of the pairing correlation on decay widths. 
T. O. thanks T. Yamashita in 
the Cyberscience Center of Tohoku University 
for a technical help for numerical calculations. 
This work was supported by the Global COE Program titled 
``Weaving Science Web beyond Particle-Matter Hierarchy'' at Tohoku 
University, and 
by a Grant-in-Aid for Scientific Research under the Program No.
(C) 22540262 
by the Japanese Ministry of Education, Culture, Sports, 
Science and Technology. 
\end{acknowledgments}

\include{end}

\end{document}